\documentclass[]{spieman} % use this instead for A4 paper
\usepackage{amsmath,amsfonts,amssymb}
\usepackage{float}
\usepackage{graphicx}
\usepackage{setspace}
\usepackage{tocloft}
\usepackage{lineno}
\usepackage{booktabs}
\usepackage{authblk}
\usepackage{geometry}
\usepackage[nohyperlinks]{acronym}
\usepackage{multicol}
\usepackage{textcomp}
\usepackage{xcolor}
\providecommand{\sisetup}[1]{}
\providecommand{\SI}[2]{\ensuremath{#1\,#2}}
\providecommand{\SIrange}[3]{\ensuremath{#1\text{--}#2\,#3}}

\providecommand{\per}{\mathbin{/}}
\providecommand{\kilo}{\mathrm{k}}
\providecommand{\mega}{\mathrm{M}}
\providecommand{\milli}{\mathrm{m}}
\providecommand{\micro}{\mu}
\providecommand{\nano}{\mathrm{n}}
\providecommand{\metre}{\mathrm{m}}
\let\meter\metre
\providecommand{\second}{\mathrm{s}}
\providecommand{\hertz}{\mathrm{Hz}}
\providecommand{\hour}{\mathrm{h}}

\title{CHARA Array Delay Lines: Upgrades, Performance and Future Directions}

\author[a*]{Narsireddy Anugu}
\author[a]{Nils H. Turner}
\author[a]{Theo A. ten Brummelaar}
\author[a]{Gail H. Schaefer}
\author[b]{Philippe B\'erio}
\author[a]{Christopher D. Farrington} 
\author[a]{Becky Flores}
\author[a]{Douglas R. Gies}
\author[c]{Stefan Kraus}
\author[a]{Edgar R. Ligon III}
\author[a]{Olli Majoinen}
\author[d]{John D. Monnier}
\author[b]{Denis Mourard}
\author[a]{Nicholas J. Scott}
\author[a]{Norman L. Vargas}

\affil[a]{The CHARA Array of Georgia State University, Mount Wilson Observatory, Mount Wilson, CA 91203, USA}
\affil[b]{Univ.\ Côte d’Azur, Observatoire de la Côte d'Azur, CNRS, France}
\affil[c]{Astrophysics Group, University of Exeter, Exeter, EX44QL, UK}
\affil[c]{Astronomy Department, University of Michigan, Ann Arbor, MI 48109, USA}

\cftpagenumbersoff{figure}
\cftpagenumbersoff{table} 
\graphicspath{{.}{Figs/}}

\begin{document} 
\maketitle

\begin{abstract}
Long baseline optical and infrared interferometric arrays achieve high angular resolution and enable detailed astrophysical measurements. Interferometers have enabled observations of stars at various stages of evolution, as well as studies of binary stars, circumstellar disks, and active galactic nuclei. The CHARA Array is a long-baseline interferometric array at the Mount Wilson Observatory, USA. At the core of CHARA operations are the delay lines, which equalize the optical path length for all telescopes as the Earth rotates and compensate for optical path variations induced by atmospheric turbulence. We report recent upgrades and performance of the CHARA Array optical delay lines for high-precision interferometric observations. The legacy system had been operational for over two decades, and it was increasingly difficult to acquire replacement parts. Beginning in mid-2021, the control system underwent a major upgrade, replacing the aging VME-based architecture with a modern hybrid FPGA and Linux-based system; this modernization continued through the end of 2024. We describe hardware/software changes, the servo architecture, and lab/on-sky performance. The upgraded system achieves residual delay line cart tracking errors of $\sim12$~nm, the same level as the legacy system, and a control bandwidth of 100-130~Hz, allowing fringe tracking across the R, H, and K bands. Initial commissioning revealed key issues such as metrology time-tick jitter and vibration-induced visibility loss, which were diagnosed and resolved. We note ongoing and future efforts to extend baselines up to 1~km and support advanced observing modes such as dual-field interferometry and nulling. This paper is a reference for current and future use of the CHARA Array and for next-generation instrument design.
\end{abstract}

% Include a list of up to six keywords after the abstract

% Include email contact information for corresponding author
{\noindent \footnotesize\textbf{*}Narsireddy Anugu, \linkable{nanugu@gsu.edu} }

\section{Introduction}
\label{sect_intro}

The CHARA Array is a long-baseline optical/infrared interferometric facility\cite{tenBrummelaar2005, Gies2024}. It has contributed to stellar and binary star astrophysics. Work includes imaging of protoplanetary disks and accretion hotspots\cite{Kraus2020, Ibrahim2023, Setterholm2025}, direct imaging of an eclipsing binary system\cite{Zhao2008,Kloppenborg2010,Baron2012}, surface mapping and spot detection on active stars\cite{Monnier2007, Monnier2012, Roettenbacher2016,Martinez2021,Parks2021,Evans2024}, and detailed imaging of convective cells and mass-loss phenomena in evolved stars\cite{Norris2021, Anugu2023, Anugu2024}. CHARA also captured the fireball expansion phase of classical nova explosions\cite{Schaefer2014,Aydi2025} and studied the dusty inner region of an Active Galactic Nucleus (AGN)\cite{Kishimoto2022}. Over 287 peer-reviewed papers use CHARA data.

\begin{figure}[ht]
\centering
\includegraphics[width=0.9\textwidth]{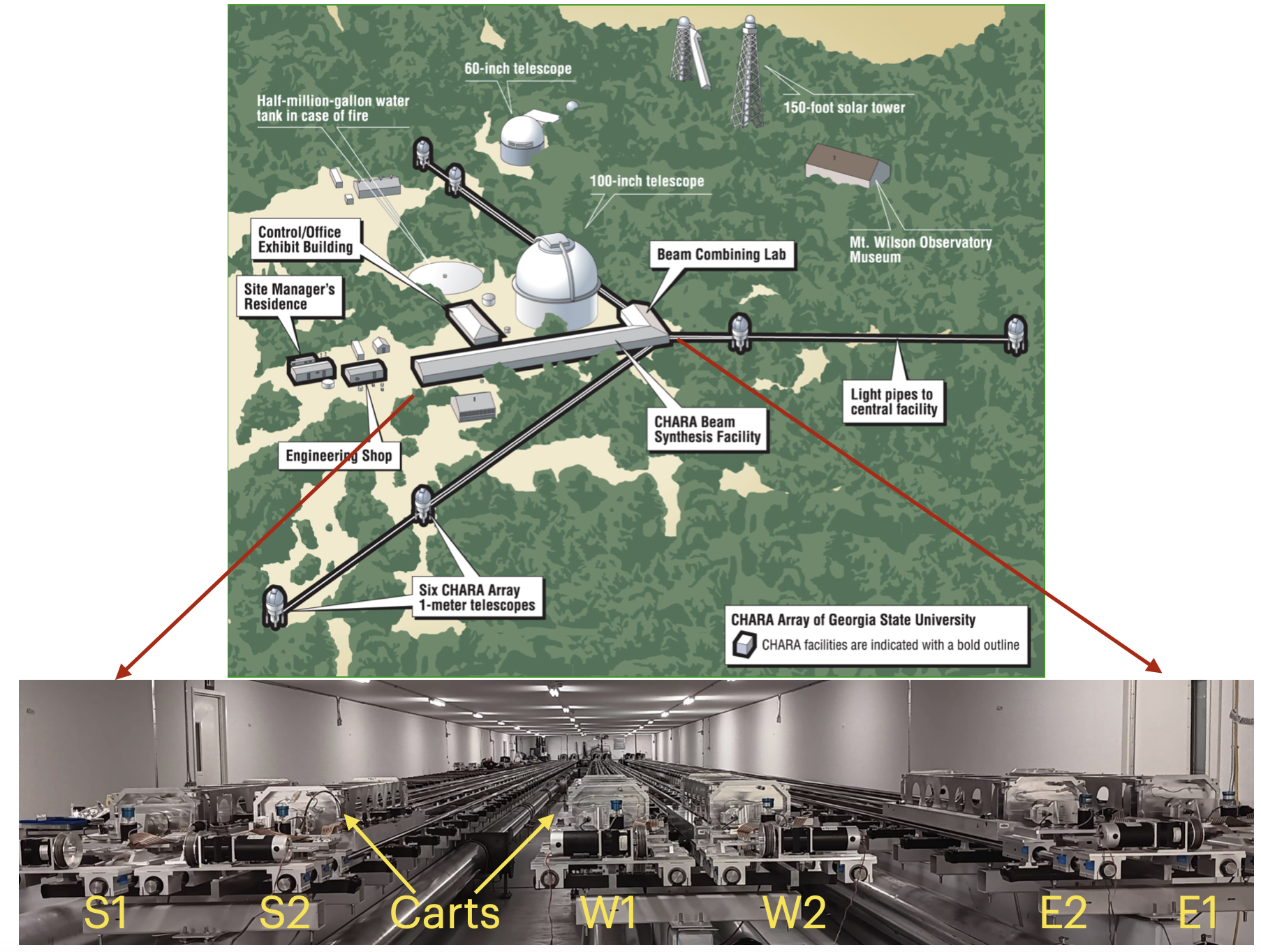}
\caption 
{\label{fig_chara} Top: Overview of the CHARA Array. The six \SI{1}{\meter} telescopes, named S1, S2, W1, W2, E1, and E2, are arranged in a Y-shaped configuration. Vacuum pipes transport starlight from the telescopes to the central beam combiner laboratory. Bottom: Photograph of the delay line system showing six optical carts that move on rails to equalize the optical path lengths for each telescope.} 
\end{figure}

The CHARA Array, operated by Georgia State University, is located at the historically significant Mount Wilson Observatory in California, USA (see Fig.~\ref{fig_chara}). As the longest baseline optical interferometer in the world, CHARA features six \SI{1}{\metre} diameter telescopes with baseline lengths ranging from $B=$~\SIrange{34}{331}{\metre}. Each telescope delivers an image with an angular resolution of $\sim\lambda/D$, where $\lambda$ is the wavelength of observation and $D$ is the telescope diameter. By coherently combining beams in the central beam combiner laboratory, the array achieves interferometric angular resolutions down to $\sim\lambda/2B$, corresponding to $\sim$\SI{200}{\micro\text{as}} resolution at $R$-band wavelengths ($\lambda \sim 0.65 \mu$m).

\begin{figure}[ht]
\centering
\includegraphics[width=0.9\textwidth]{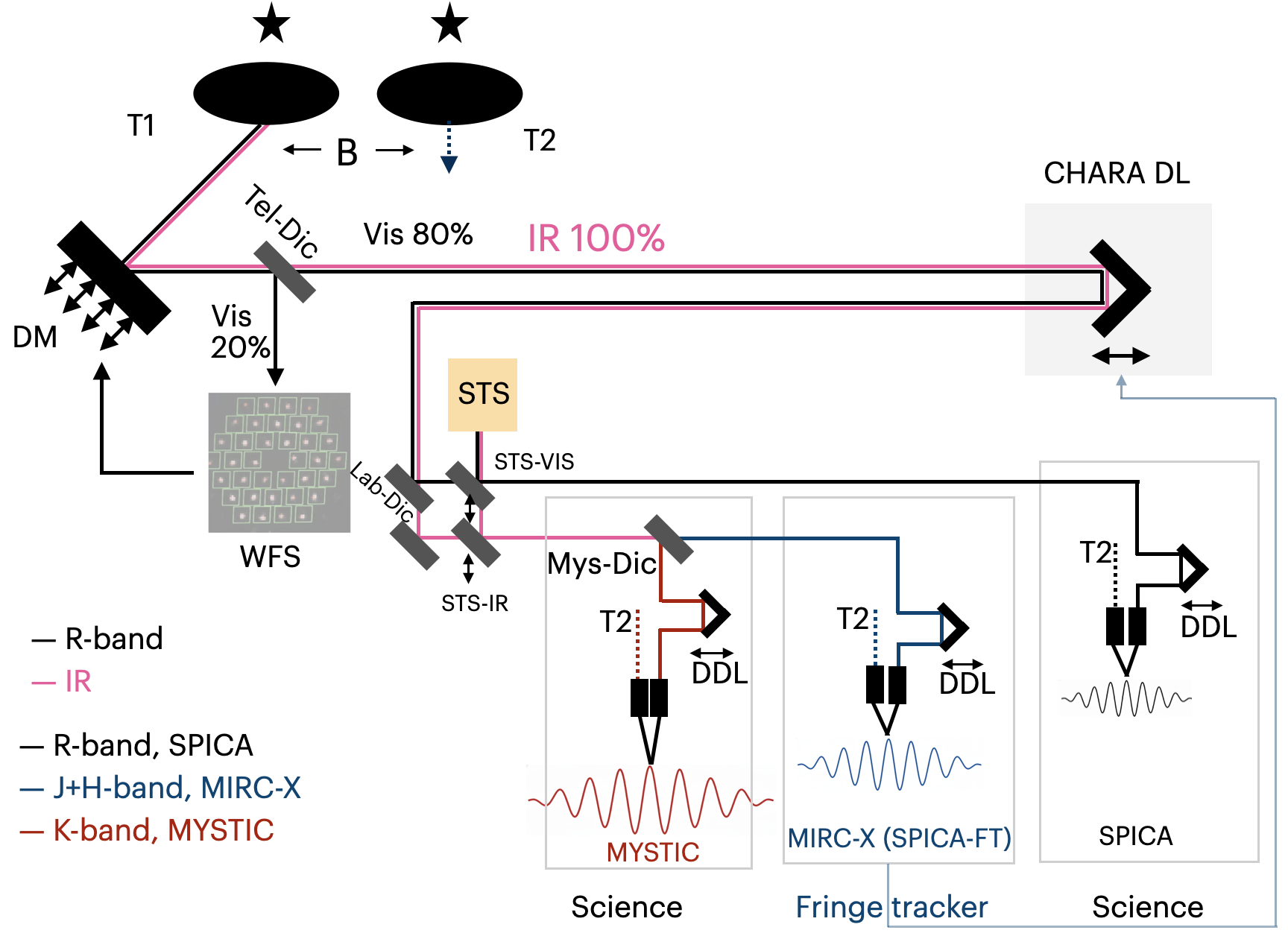}
\caption
{ \label{fig_spica_mircx_mystic_schematic}
Schematic of the CHARA beam train feeding the SPICA (visible), MIRC-X (near-IR; SPICA-FT), and MYSTIC (near-IR) beam combiners. Only one telescope is shown. In the SPICA+MIRC-X+MYSTIC mode, the telescope dichroic (TelDic) directs 20\% of the visible light to the AO/WFS and transmits the remaining visible light together with the near-IR beam to the beam-combiner laboratory. After the common CHARA delay lines, laboratory dichroics (Lab-dic)  separate the visible and near-IR beams to the SPICA and MIRC-X/MYSTIC beam-combiner tables. MIRC-X provides fringe-tracking corrections (SPICA-FT) applied through the common delay lines, while SPICA and MYSTIC record science data; internal delay stages compensate chromatic dispersion and non-common-path optical path differences. A longitudinal dispersion corrector located between the delay lines and the laboratory dichroics is not shown. The diagram shows the functional architecture; the full six-telescope lab layout is omitted.
}
\end{figure}

A typical CHARA observation begins with acquiring starlight at each telescope, followed by correction of atmospheric wavefront distortions by the adaptive optics system (AO) \cite{Che2013, Anugu2020b}. All acronyms are defined in Appendix~\ref{sec:acronyms}. The AO-corrected beams are then transported using vacuum pipes from each telescope to the beam combiner laboratory, where delay lines equalize the optical path lengths (see Fig.~\ref{fig_chara} and \ref{fig_spica_mircx_mystic_schematic}). The beams are then injected into the selected beam combiner instruments to record interference fringes. The CHARA Array is equipped with three six-telescope beam combiners that support simultaneous observations across a broad wavelength range (0.63 -- 2.4~\textmu m), covering the $R$, $I$, $J$, $H$, and $K$-bands. These instruments -- (i) Stellar Parameters and Images with a Cophased Array \cite{Mourard2024} (SPICA, $R/I$ bands, 0.63-0.95~\textmu m), (ii) Michigan InfraRed Combiner-eXeter \cite{Anugu2020} (MIRC-X, $J/H$ bands, 1.1-1.8 \textmu m), and (iii) Michigan Young Stellar Imager at CHARA \cite{Setterholm2023} (MYSTIC, $K$-band, 2.0-2.4 \textmu m) – each provide squared visibilities on 15 baselines and closure phases on 10 unique triangles, enabling high-fidelity imaging. Imaging at CHARA was concentrated in the near-infrared; with the installation of SPICA, simultaneous imaging across $R$- to $K$-bands is now possible. CHARA includes other combiners such as CLASSIC \cite{tenBrummelaar2013} ($J$, $H$, $K$-bands), Silmaril \cite{Lanthermann2024} ($H+K$ bands) and the CHARA Array Integrated Optics Test bench \cite{Mayer2024} (CHARIOT, $K$-band), which support two- or three-telescope observations with a focus on sensitivity and testing novel astro-photonics-based technologies.

%Add the schematic figure here

The delay line system of CHARA (see Sec.~\ref{sec_overview_delaylines}), developed in the early 1990s, relied on a VME-based control architecture with PowerPC processors running the VxWorks real-time operating system. All six delay lines shared the same VxWorks control system in the legacy setup. While the legacy system served reliably for over 20 years, it became increasingly difficult to acquire replacement parts. On a few occasions, CHARA operations had to be temporarily suspended while components were shipped to the vendor for repair and returned, which caused downtime. In August 2021, the delay line system underwent a modernization. A hybrid FPGA--Linux architecture replaced the legacy system to improve maintainability and support fast fringe-tracking communications (Sec.~\ref{sec_architecture}). The core hardware/FPGA platform delivered and installed by AZ Embedded Systems; the CHARA team later found issues and troubleshot the installed system.

After deployment, the new FPGA based delay line control system initially exhibited several performance-limiting issues during early operations. Early issues were dominated by timing-jitter-induced tracking errors and their diagnosis and mitigation are detailed in Sec.~\ref{sec_initial_results}. The cart tracking errors $\epsilon(t) = P_{\mathrm{target}}(t) - P_{\mathrm{measured}}(t) \sim \SI{80}{\nano\metre}$ RMS were caused primarily by timing inaccuracies within the metrology FPGA and by suboptimal gain tuning in the control loop, which initially constrained the system bandwidth and impaired the responsiveness necessary for real-time fringe tracking. Here, $P_{\mathrm{target}}(t)$ is the commanded target position, $P_{\mathrm{measured}}(t)$ is the actual position of the cart measured by the metrology laser, and $t$ is time. These issues were especially detrimental for fringe detection in the visible-light ($R$-band) beam combiners that rely on precise delay corrections. Later software fixes reduced these issues, improving tracking accuracy and delay line performance.

Although CHARA has been in routine operation since 2003, detailed documentation of its delay line performance — especially post-upgrade —remains limited. We describe the implementation and characterize the upgraded system to help interpret CHARA observations and guide future instrument design. 

The paper is organized as follows: Sec.~\ref{sec_tech_require} provides technical requirements. Sec.~\ref{sec_overview_delaylines} describes the delay line hardware and control system. Sec.~\ref{sec_architecture} outlines the overall system architecture, encompassing both hardware and software components. Sec.~\ref{sec_initial_results} reviews key lessons learned during the initial commissioning phase in mid-2021. Sec.~\ref{sec_results} presents performance results from laboratory measurements and on-sky observations. Finally, Sec.~\ref{sec_future} discusses ongoing developments and planned future upgrades.

%%%%%%%%%%%%%%%%%%%%%%%%%%%%%%%%
\section{Technical Requirements}\label{sec_tech_require}
This section lists technical performance requirements for the CHARA delay lines needed for scientific operations. High instrument fringe visibility is essential for interferometric observations and depends on the precision and responsiveness of the delay line system. Fringe visibility quantifies the coherence of the interference pattern, where a visibility of unity indicates ideal interference—marked by fully constructive and destructive fringes—while a visibility of zero corresponds to complete incoherence and no detectable fringe signal.

On-sky system visibility is influenced by a variety of factors \cite{Eisenhauer2023,Anugu2020,Setterholm2023,Mourard2024}, including residual delay line tracking errors, the intrinsic structure of the astrophysical source, atmospheric turbulence, wavelength dispersion \cite{Pannetier2021}, polarization mismatches, flux imbalance between telescopes, mechanical vibrations, and non-common path optical path difference (NCOPD) residuals between the fringe tracker and the science instrument (see Fig.~\ref{fig_spica_mircx_mystic_schematic}). Among these, inaccurate delay line tracking is often a dominant source of fringe visibility degradation.

Given that residual wavelength dispersion and polarization mismatches alone can contribute more than 5\% visibility loss \cite{Pannetier2021,Anugu2020}, we conservatively allocate a maximum of 5\% visibility loss to the delay line system. To maintain $\geq95$\% fringe contrast in CHARA science observations — especially in the $R$, $H$, and $K$-bands — the delay lines must achieve cart tracking errors of $\epsilon(t) \le \SI{20}{\nano\metre}$ and operate with response times faster than the atmospheric coherence time, typically $\tau_0 = \SIrange{6}{30}{\milli\second}$. See Table~\ref{tab:tracking_require} for expected tracking error tolerances across wavelength bands, derived from Eq.~\ref{v_loss} and the visibility contrast model of Conan et al. (1995) \cite{Conan1995}.

\begin{table}
\centering
\caption{CHARA delay line tracking accuracy requirements, derived to maintain fringe visibility contrast loss $\leq5\%$ across the $R$, $J$, $H$, and $K$-bands, based on the visibility contrast model in Eq.~\ref{v_loss} \cite{Conan1995}. Atmospheric coherence times ($\tau_0 = r_0/v$) are estimated to lie between \SIrange{1}{7}{\milli\second} at $\lambda=$\SI{443}{\nano\metre}, based on empirical measurements at Mount Wilson Observatory \cite{Buscher1994}, where $v$ is the wind speed. These measurements are consistent with our new seeing monitor built by Univ.\ Côte d’Azur (Mourard et al. in prep). The $\tau_0$ values for each band are computed using the wavelength scaling $\tau_0 \propto \lambda^{6/5}$, with central wavelengths representative of each photometric band.}
\begin{tabular}{lcccc}
\toprule
\textbf{Parameter} & \textbf{$R$-band} & \textbf{$J$-band} & \textbf{$H$-band} & \textbf{$K$-band} \\
\midrule
Central Wavelength [$\lambda_0 = $\textmu m]      & 0.630 & 1.25 & 1.65 & 2.20 \\
Cart tracking requirement [nm RMS]     & 20  & 40  & 50  & 75  \\
Mean $\tau_0$ [$\lambda^{6/5}$ scaled] [ms] & 6.1  & 13.9 & 19.4 & 27.3 \\
Best-case $\tau_0$ [ms]        & 11.0 & 24.3 & 34.0 & 48.0 \\
\bottomrule
\end{tabular}
\label{tab:tracking_require}
\end{table}

\begin{equation}\label{v_loss}
  V \sim \exp\left[-\left(\frac{2\pi}{\lambda} \epsilon \right)^2\right]
\end{equation}

The delay line optical path difference (OPD) equalization for starlight can be broadly categorized into two main components:

\begin{itemize}
\item Predictable delays are geometric delays caused by the telescope geometric configuration, sidereal rotation of Earth, and differential air paths between telescopes. These delays are modeled as a ``baseline solution" and provided to the delay line metrology system as continuous target positions. Furthermore, delays introduced by the differential air paths between the telescope arms to the beam combiner laboratory \cite{Pannetier2021}. Executing this baseline solution requires the delay line carts to move at speeds of up to \SI{8}{\milli\metre\per\second}.
\item Dynamic and fast-changing $\Delta$OPDs result from atmospheric turbulence and mechanical vibrations. These fluctuations are measured in real time by fringe trackers and actively corrected via feedback to the delay lines. Fringe tracking compensates for the phase delay variations introduced by the turbulent atmosphere and must operate faster than the atmospheric coherence time $\tau_0$ (see Table~\ref{tab:tracking_require}).
\end{itemize}

To meet these demands, the delay line system must fulfill two key technical requirements:

\begin{itemize}
  \item Accuracy: The delay line carts should track with residuals $\epsilon(t)\leq\SI{20}{\nano\metre}$ RMS for the fringe detection with $\geq95\%$ instrument fringe contrast.
  \item Speed: It must correct quickly enough—within \SI{6}{\milli\second} to correct for fast fluctuations caused by the atmosphere.
\end{itemize}

CHARA uses two types of fringe tracking systems to manage these dynamic OPD fluctuations:

\begin{itemize}
  \item Group-delay (fringe-packet) tracking uses spectrally dispersed fringes (typically spanning 10--20 wavelengths) to estimate atmospheric $\Delta$OPDs within the spectral coherence length ($\tfrac{\lambda^2}{\Delta \lambda}$), which scales with spectral resolution ($\mathcal{R} \propto \tfrac{\lambda}{\Delta \lambda}$). This approach supports slow-timescale tracking (often at $\sim$\SI{1}{\hertz}) and can accommodate larger OPD offsets because it operates under more relaxed phase-stability requirements. At CHARA, MIRC-X and MYSTIC implement group-delay tracking.
  \item Phase tracking applies real-time OPD corrections on sub-wavelength scales ($<\lambda$) at high loop rates (typically $\gtrsim$\SI{100}{\hertz}), enabling coherent integration over longer exposures. At CHARA, SPICA-Fringe Tracker (SPICA-FT) \cite{Pannetier2022,Mourard2024}, a software application built on the MIRC-X hardware and software platform, operates at \SIrange{250}{500}{\hertz} and achieves \SIrange{90}{230}{\nano\metre} RMS residual OPD (Sec.~\ref{sec_results}). While developed primarily to enable longer coherent integrations for SPICA, this capability also benefits other CHARA beam combiners.
\end{itemize}

These requirements — RMS cart tracking error of $\epsilon(t) \leq \SI{20}{\nano\metre}$ and correction speeds faster than $\tau_0 = \SI{6}{\milli\second}$ — serve as the benchmark criteria used throughout this paper to evaluate the success of the upgraded system.

\begin{figure}
\centering
\includegraphics[width=0.9\textwidth]{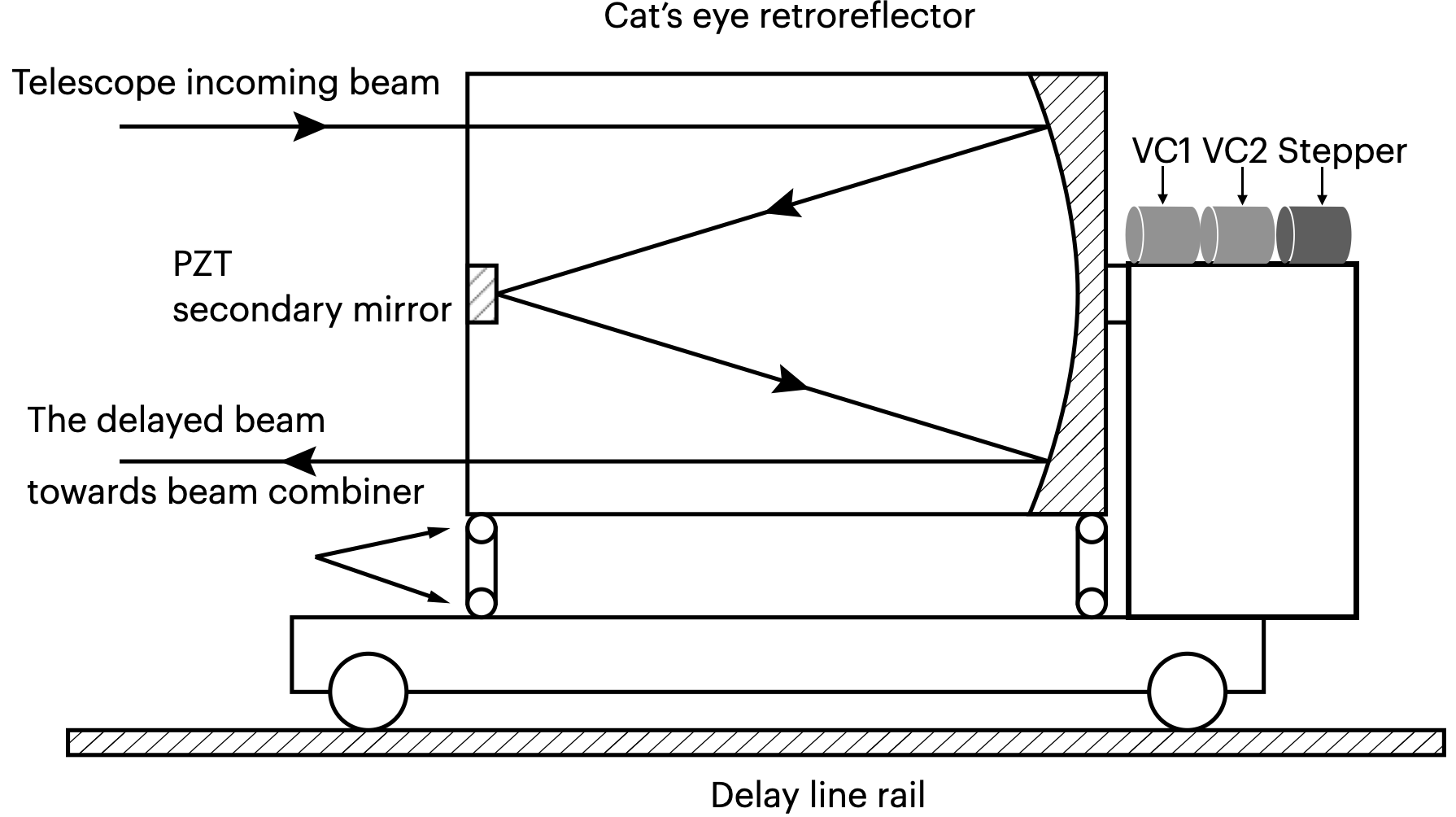}
\caption 
{ \label{fig_delayline_cart}
Schematic diagram of a delay line cart with a cat's eye retro-reflector used at CHARA. The telescope incoming beam enters the cat's eye system and reflects off one side of a 
paraboloid mirror and then focuses onto a small secondary flat mirror, which redirects the beam back through the second side of paraboloid. The result is a re-collimated output beam traveling parallel to the input. The secondary flat mirror is mounted on a high-bandwidth PZT stage to enable fine optical path-length adjustments. The PZT, Voice Coil 1 and Voice Coil 2 and stepper motor work in a cascaded servo system.
}
\end{figure} 

\begin{figure}[ht]
\centering
\includegraphics[width=0.95\textwidth]{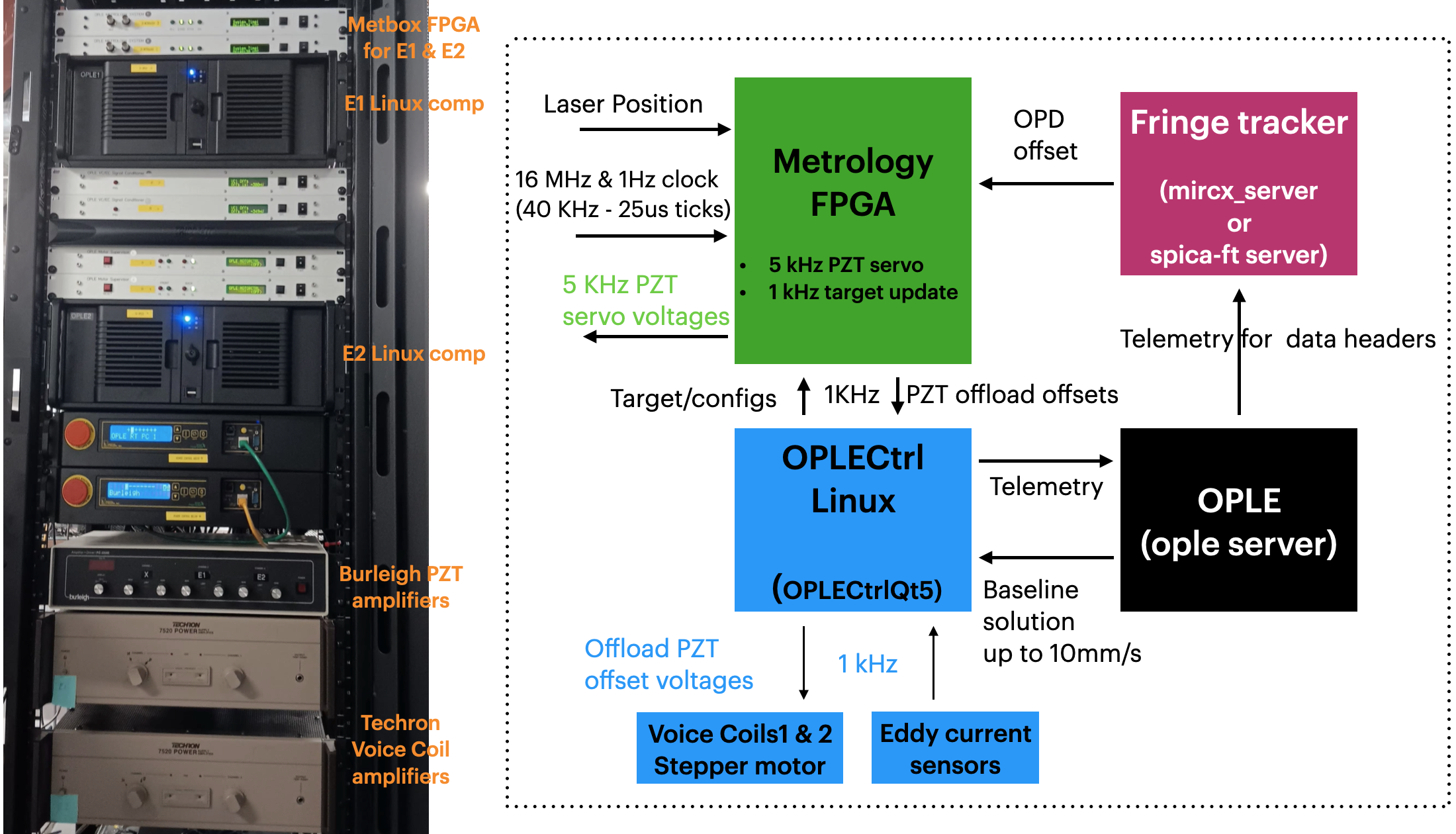}
\caption 
{ \label{fig_architecture} CHARA OPLE control system architecture and hardware layout. The left panel shows the physical hardware rack for the E1 and E2 delay lines, which includes FPGA-based metrology systems, Linux control computers, and PZT and voice coil voltage amplifiers (see~Sec.~\ref{sec_architecture}). The right panel presents the control logic. At the core of the system is the time critical Metrology FPGA, which receives absolute laser-based position measurements and high-precision clock signals \cite{Sturmann1998} (\SI{16}{\mega\hertz} and \SI{1}{\hertz}). It operates two control loops: (i) a \SI{5}{\kilo\hertz} PZT servo loop, and (ii) a \SI{1}{\kilo\hertz} loop that handles slower tasks such as target position updates and the transmission of telemetry data—including PZT positions and cart positions—to the Linux computer. The soft real-time Linux server (running \texttt{OPLECtrlQt5}) acts as the supervisor and controls the voice coils and stepper motors based on PZT offloads and eddy current sensors — at \SI{1}{\kilo\hertz}. The \texttt{ople server} and \texttt{opletab} graphical user interface, operated by the CHARA telescope operator, handle cart selection, homing, baseline solution activation, and monitoring. The \texttt{ople server} sends baseline solution model, accounting for sidereal motion and array geometry, which are applied to the delay line motion at speeds of up to \SI{8}{\milli\metre\per\second}. The fringe tracker (e.g., SPICA-FT, MIRC-X, or MYSTIC) sends atmospheric residual OPD offsets directly to the metrology FPGA, which are corrected by the PZT servo. } 
\end{figure}

%%%%%%%%%%%%%%%%%%%
\section{Overview of Delay Lines: Opto-Mechanical Systems}\label{sec_overview_delaylines}

The six CHARA delay lines are designated as OPLE (Optical Path Length Equalizers). Each OPLE consists of a precision-controlled cart that carries a retro-reflecting cat's eye optical assembly along a \SI{46}{\meter} rail track, providing \SI{92}{\meter} of optical delay (twice the physical delay) \cite{tenBrummelaar2005}. During on-sky operations, one delay line cart is assigned as the reference, and the remaining five carts move with respect to the reference cart.

The overall design of the delay lines is based on the architecture pioneered at JPL~\cite{Shao1988,Colavita1991,Colavita1992}, and subsequently deployed in the Palomar Testbed Interferometer (PTI)\cite{Colavita1999} and the Keck Interferometer (KI)\cite{Colavita2013}. Each cart system comprises five components: (i) optical assemblies for delay implementation, (ii) sensors to measure cart positions, including metrology laser and eddy current sensors, (iii) actuators for fine and coarse motion control, and (iv) a high-bandwidth control computer system. The design considerations for mechanical support structures \cite{Ridgway1998Design,Ridgway1997TSupport,Barr1997Drive,tenBrummelaar1996OPLE} and optical mirrors \cite{tenBrummelaar1996Mirror,Sturmann2021Alignment} are well documented in the CHARA technical report series.

The cat's eye \cite{Shao1988,Colavita1991} optical assembly implements the optical delay as follows: light enters the system, reflects off one side of a $13\times8$~inch paraboloid mirror (42-inch focal length), comes to a focus on a small secondary flat mirror, and is redirected back through the paraboloid to emerge as a re-collimated beam traveling parallel to the input beam. The secondary flat is mounted on a high-bandwidth Piezoelectric Transducer (PZT) stage to enable fine path-length adjustments.

The optical assembly is supported by flexure arms mounted to the optics cart, which travels on stainless steel wheels along the rail system. The optics assembly can move independently from the cart via a voice coil for intermediate-scale motion control. A separate motor cart, driven by a friction-coupled micro stepper, provides coarse translation by pushing the optics cart. A cable management system, including an idler pulley and a torque-motor-tensioned ribbon cable, delivers power and control signals to all mobile elements.

Each cart motion control is implemented in four stages with four complementary actuators: a PZT for high-speed correction, an optics voice coil, a stepper voice coil to correct the PZT offloads, and a stepper motor for coarse cart positioning (see Fig.~\ref{fig_delayline_cart}). 

The cart position feedback is provided by three complementary sensors: a 1.319~\textmu m heterodyne laser metrology system, an eddy current sensor on the optics cart, and an eddy current sensor on the stepper system. The tracking servo system is a cascaded design, where the input of each servo system depends on the output of the previous system.

The PZT servo system attempts to correct the metrology laser error, $P_{\mathrm{target}}(t) - P_{\mathrm{measured}}(t)$. The optics voice coil system (VC1) uses the piezo position as error input and adjusts the optics voice coil to keep the PZT near its center position. The stepper voice coil system (VC2) uses the position of the optics cart (eddy current sensor 1) as error input and adjusts the stepper voice coil to keep the optics cart position centered. Finally, the stepper servo system uses the stepper cart position (eddy current sensor 2) as error input and actuates the stepper motor to keep the stepper cart position centered.

At the core of the delay line system is the precise measurement of the cart position with nanometer-level accuracy. This is accomplished using a polarization heterodyne laser interferometer\cite{Shao1988,Colavita1991,Colavita1992,Colavita1999}. A stabilized heterodyne metrology laser operating at \SI{1.319}{\micro\metre} serves as the light source. Each metrology laser head uses a polarizing beam splitter to separate the beam into two orthogonal polarization components: one acts as a stable phase reference, while the other is used to measure the optical path length in a double-pass configuration. These two beams are frequency-shifted to create a beat frequency with a \SI{2}{\mega\hertz} difference. The phase of the resulting \SI{2}{\mega\hertz} signal from the heterodyne interferometer is compared to the \SI{2}{\mega\hertz} reference signal, enabling precise measurement of the optical path delay. The resulting metrology signals are captured by a custom multichannel fringe-counter card and transmitted via optical fibers to the metrology FPGA system (Sec.~\ref{sec_architecture}) for real-time processing. This heterodyne fringe-counting approach provides an absolute, unwrapped path-length measurement after homing, rather than an incremental encoder measurement.

\begin{figure}
\centering
\includegraphics[width=0.85\textwidth]{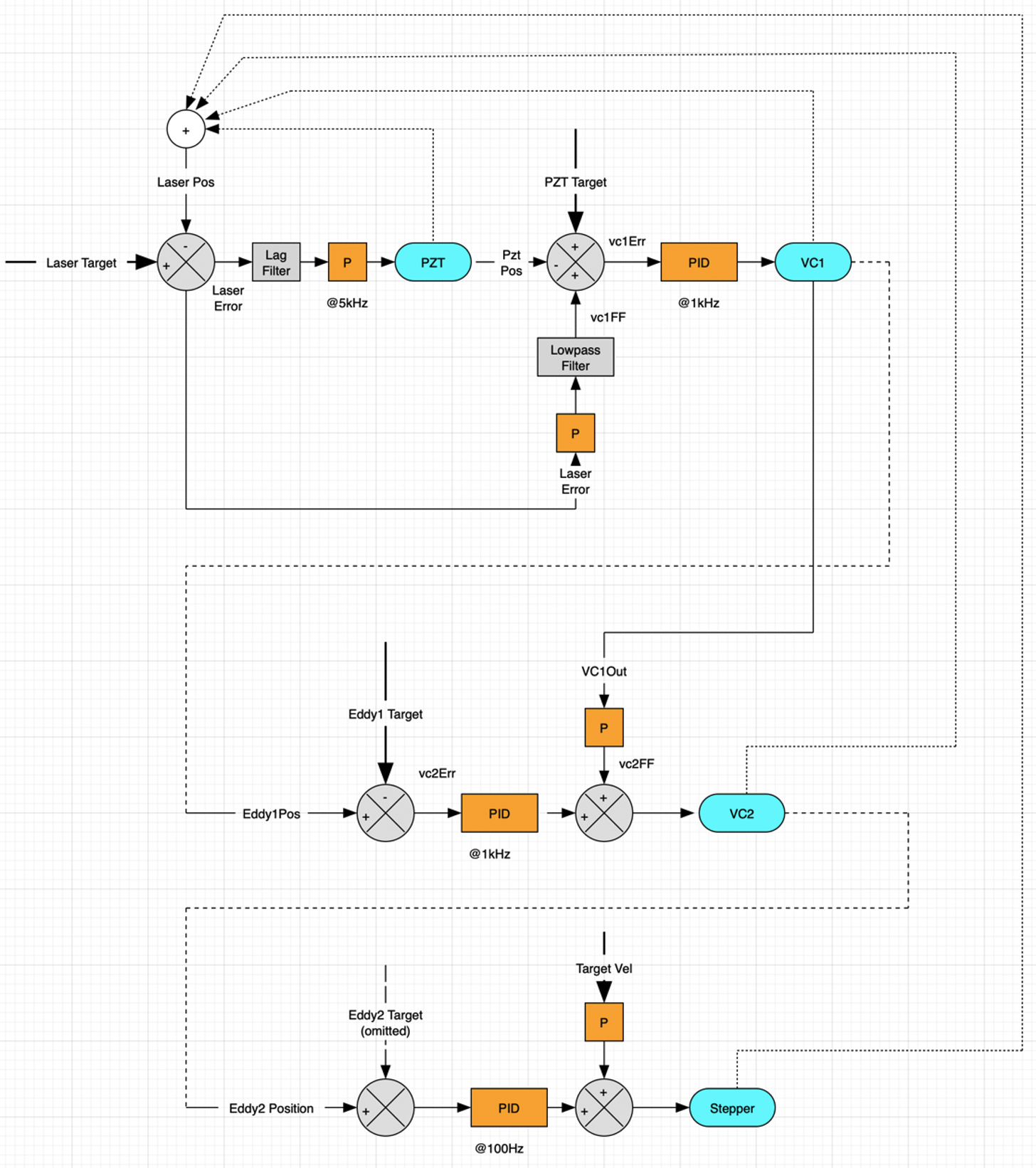}
\caption 
{ \label{fig_servo_tracking}
Nested (cascaded) OPLE tracking servos. The metrology error drives the PZT loop (fast), which is offloaded sequentially to VC1 (optics cart centering), VC2 (stepper interface centering), and the stepper motor (coarse positioning). See Sec.~\ref{sec_architecture} for details. } 
\end{figure}

%%%%%%%%%%%%%%%%%%%%%%%%%%%%%%%%%%%%%
\section{Overview of Delay Lines: Control Hardware \& Software Architecture}\label{sec_architecture}
Fig.~\ref{fig_architecture} presents the upgraded control system architecture. Replacing the legacy VxWorks real-time operating system, the new control system implements a hybrid architecture, combining FPGA-based real-time control with a Linux-based soft real-time backend. Each delay line cart is now controlled by a dedicated Linux workstation, improving modularity and fault isolation. The system is fully integrated into the CHARA messaging infrastructure, eliminating legacy software dependencies and enhancing maintainability.

\subsection{Embedded Metrology FPGA System + PZT Actuator Servo}
The OPLE embedded metrology system is the most timing-critical component of the delay line control system. It is implemented using a 32-bit NIOS soft-core processor, clocked at \SI{100}{\mega\hertz}, on an Altera DE2-115 development board featuring a Cyclone IV FPGA. The system is developed using the Intel Quartus 18.1 development platform and handles the most timing-critical scheduling of the control system.

The embedded metrology system receives 1.319~\textmu m metrology laser signals via optical fibers and timing synchronization clocks (\SI{1}{\hertz} and \SI{16}{\mega\hertz}) from the CHARA Master Clock installed on the GPS computer. The system digitizes the metrology laser signals into square waveforms and converts them into position measurements. Based on these position measurements, it generates new targets and implements servo control for the PZT actuator mounted on the optics cart. This system is also responsible for target calculations based on star baseline solution and fringe tracking offsets (time, offset position, velocity). The metrology signal acquisition and PZT servo control operate within a \SI{5}{\kilo\hertz} loop. The metrology system sends status and position updates at \SI{1}{\kilo\hertz} to the Linux control computer using User Datagram Protocol (UDP) communication protocol, then it triggers the movement of other, less timing-critical servo systems, such as voice coils and stepper motors.

\subsection{Voice Coil Actuators}
The OPLE cart is equipped with two voice coil actuators and two eddy current sensor systems, which actuate the optics cart assembly and the interface between the optics and stepper cart. The voice coil systems are actuated using Techron voice coil amplifiers controlled by an input signal ranging from +10V to -10V into a high-impedance input. The eddy current sensor system measures the position of the voice coils and their effect on the related cart system. Two voice-coil stages combine high-bandwidth, small-range correction (VC1) with a larger-range centering stage (VC2) to avoid saturating the fast stage while preserving bandwidth.

\subsection{Stepper Motor Controller}
A Parker LN stepper driver executes micro-stepping at 50 k steps/revolution and receives motion commands from the Linux computer at \SI{100}{\hertz}.

\subsection{Top-Level Linux Control}
The control computers operate on a low-latency Ubuntu 20.04 LTS distribution. These computers, one for each cart, house Analog-to-Digital Converter / Digital-to-Analog Converter cards and a motion controller (MFX1040) to move the voice coils and stepper actuators. The core \texttt{OPLECtrlQt5} software, built using Qt cross-platform software, manages the servo loops and provides a Graphical User Interface (GUI) for real-time monitoring and control.

\subsection{Control System Software}
The metrology FPGA system updates the target position at \SI{1}{\kilo\hertz} and produces smooth motion trajectories. In tracking mode, four nested servo loops are implemented (see Fig.~\ref{fig_servo_tracking}). The PZT servo loop runs within the embedded metrology FPGA system. The Voice Coil 1 and 2 and stepper motor loops run in Linux machine with server \texttt{OPLECtrlQt5}.

\begin{enumerate}
\item PZT Servo Loop (\SI{5}{\kilo\hertz}): A fast Proportional, Integral, Derivative (PID) servo loop runs on the laser metrology error. It uses proportional control and lag compensation to actuate the PZT with high stability. The Burleigh amplifier supplies the required high voltage (0--120V~DC).
\item Voice Coil 1 Loop (\SI{1}{\kilo\hertz}): It keeps the PZT actuator centered (-32 to 32 \textmu m optical delay range peak-to-peak) using the piezo position as feedback. It also includes a feedforward term from laser error. This voice coil has 6 mm range.
\item Voice Coil 2 Loop (\SI{1}{\kilo\hertz}): It uses Voice Coil 1 (optics cart position) as feedback to maintain centering. It adds feedforward from Voice Coil 1 to prevent saturation. PID loop using eddy current sensor 1. This voice coil has 12 mm range.
\item Stepper Motor Loop (\SI{100}{\hertz}): Uses eddy current sensor 2 for error feedback and adjusts the motor rate accordingly. A Proportional-Derivative (PD) servo loop maintains centering using eddy current sensor 2.
\end{enumerate}

During large slews, the controller switches to a special mode that disables the PZT loop and relies on voice coils and motor drives. The modular design allows individual testing of each loop and supports fault-tolerant operation across the control architecture.
PID gain tuning is done in stages: the PZT loop is tuned first for stability and bandwidth, then VC1 and VC2 are tuned to keep the PZT centered while maintaining adequate stability margins, and finally the motor loop is tuned for slow centering; derivative terms use a low-pass (lag) filter to limit noise amplification.

%-----------------------------
\subsection{On-sky operations}

For operators, many technical details are abstracted during observations. Typically, observers interact with the top-level software servers and GUIs. Two key servers are involved in operations: one is managed by the CHARA telescope operator (\texttt{ople server} and \texttt{opletab} GUI), and the other by the fringe tracker instrument observer (e.g., MIRC-X, MYSTIC, or SPICA-FT). This section uses SPICA-FT as an example. Three computers are involved: OPLECtrl (running \texttt{OPLECtrlQt5}), OPLE (running \texttt{ople server}), and MIRC-X (running \texttt{spica-ft server}).
The \texttt{ople server}, running on the OPLE Linux machine, manages delay line operations (see Fig.~\ref{fig_architecture}). Its responsibilities include selecting the appropriate carts, executing homing procedures, configuring fringe tracking modes, activating the baseline solution based on the star coordinates and time, and initiating tracking. These commands are transmitted to the \texttt{OPLECtrlQt5} application. As part of the star acquisition sequence, the telescope operator activates the baseline solution model, which calculates the required cart positions based on the array geometry and Earth sidereal motion and moves them accordingly. Meanwhile, the instrument observer manages the fringe tracker, scanning for fringes and applying corrections to the residual baseline solution offsets. Once the fringes are detected, the SPICA-FT server locks onto them and begins correcting for atmospheric turbulence and mechanical vibrations in real time.

When actions are triggered through the \texttt{ople server} using its client GUI (\texttt{opletab}), the commands are sent to the underlying Linux server, \texttt{OPLECtrlQt5}. This supervisor server remains invisible to astronomers but is responsible for executing the control logic initiated by the OPLE interface. Both the \texttt{ople server} and \texttt{OPLECtrlQt5} follow the CHARA-compliant architecture for communication, based on a client-server model.
SPICA-FT, which runs on the MIRC-X computer, sends its fringe tracking corrections directly to the embedded metrology system using a dedicated Ethernet connection. This architecture minimizes communication latency. Although it is technically possible to route corrections through the \texttt{ople server}, doing so introduces additional communication layers, specifically, from SPICA-FT to \texttt{ople server}, from \texttt{ople server} to \texttt{OPLECtrlQt5}, and finally to the metrology system, resulting in increased latency.

Timing is critical in the OPLE control system. Historically, the CHARA Array location allowed a simplified timing approach. Since observations occur during local nighttime, which falls within a single UTC day, time could be defined relative to midnight UTC without wraparound issues.
Time is defined in the same way as in the legacy system: as the number of \SI{25}{\micro\second} intervals (“ticks”) since midnight UTC. This corresponds to a \SI{40}{\kilo\hertz} time base derived from a shared \SI{16}{\mega\hertz} Master Clock\cite{Sturmann1998}.
All six embedded metrology systems use this common clock, ensuring synchronization with the OPLE system for coherent delay tracking. Server machines, including OPLECtrl, OPLE, and MIRC-X, are synchronized using the Network Time Protocol (NTP) to align the control framework.

\subsection{Telemetry}\label{sec_telemetry}
\begin{figure}
\centering
\includegraphics[width=0.85\textwidth]{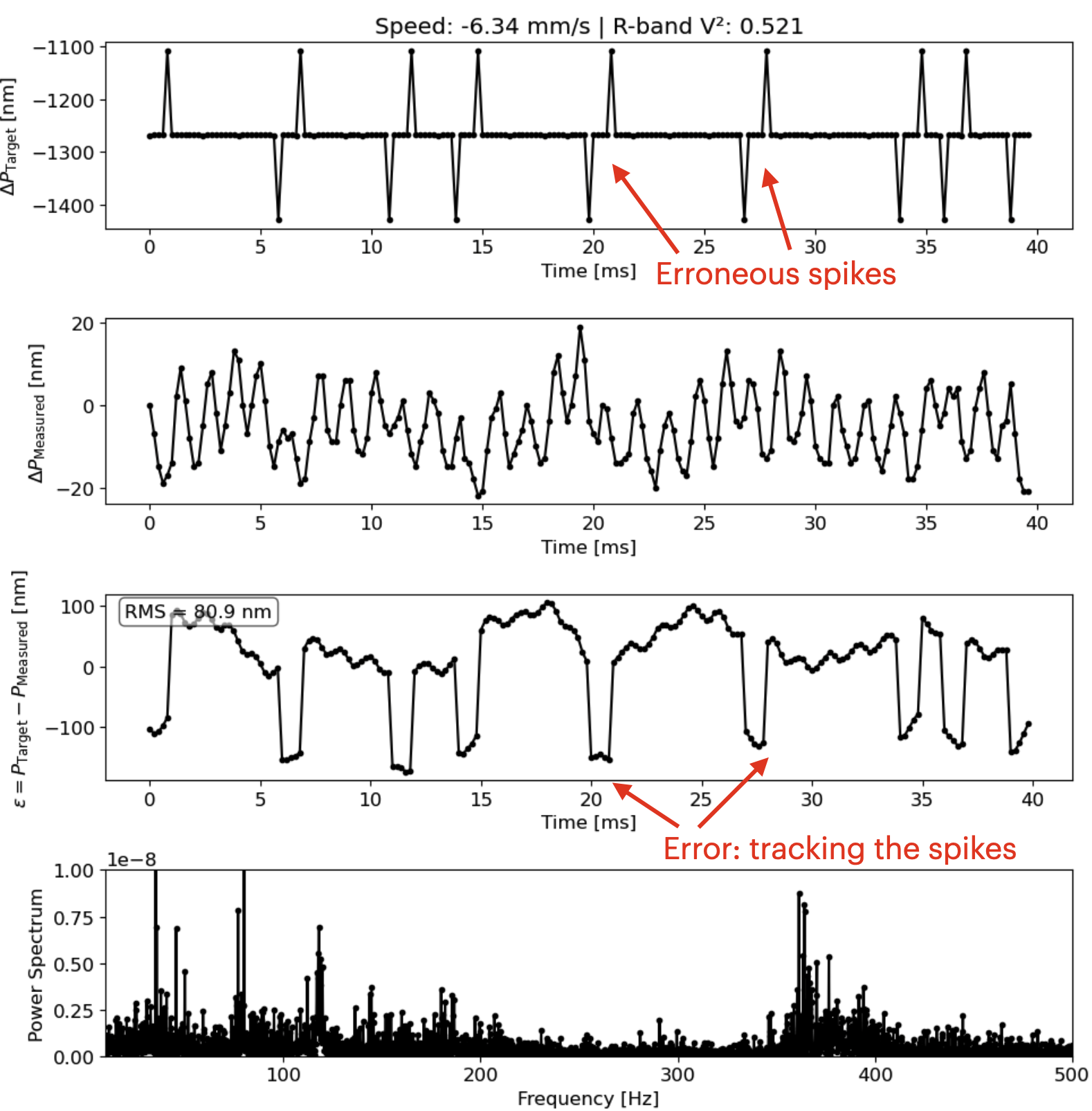}
\caption 
{ \label{fig_target_spikes} 
Example data illustrating erroneous target position spikes caused by timing jitter in the internal clock of the metrology system. Here the E1 cart was following a ``baseline solution", moving at \SI{5.52}{\milli\metre\per\second} resulting in a tracking error of approximately $\epsilon(t) \sim \SI{80}{\nano\metre}$ RMS. The fringe tracker was not sending OPD offsets. As the PZT servo attempted to track these artificial spikes, the cart experienced induced vibrations. This issue was ultimately resolved through a software update that corrected the underlying clock instability. 
} 
\end{figure}

\begin{figure}
\centering
\includegraphics[width=0.85\textwidth]{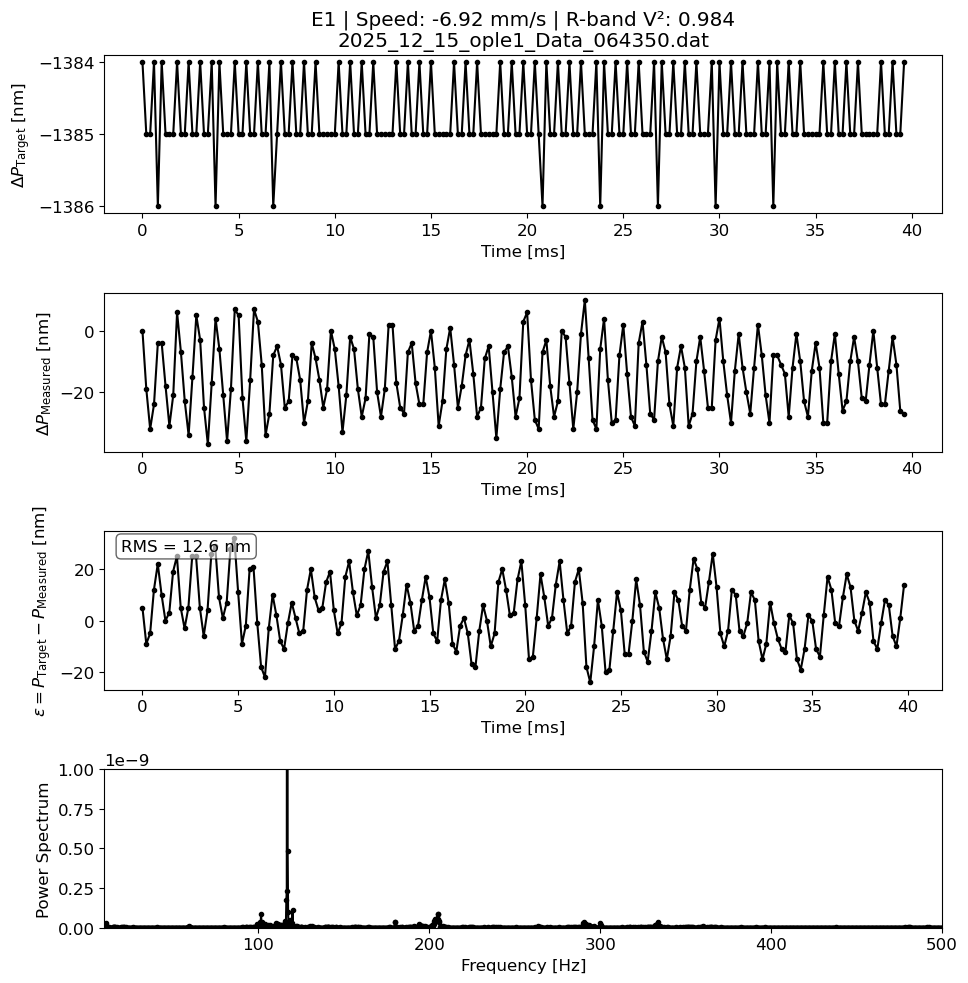}
\caption
{\label{fig_target_spikes_fixed}
Same configuration as Fig.~\ref{fig_target_spikes} after correcting the metrology clock jitter. The fringe tracker offsets were disabled and only the baseline solution (\SI{-6.92}{\milli\metre\per\second}) was applied. The tracking error is reduced to $\epsilon(t)\sim\SI{12.6}{\nano\metre}$ RMS, corresponding to $V^2\sim98\%$.}
\end{figure}

During on-sky observations, telemetry files are automatically recorded for each of the science and calibrator star sequences at a frame rate of \SI{5}{\kilo\hertz} for a 5-sec duration for each cart. These files capture raw data including the $P_{\mathrm{target}}(t)$, $P_{\mathrm{measured}}(t)$, $\epsilon(t)$, and timestamps $t$. In addition, the system logs the velocities of each actuator—such as the piezoelectric transducers (PZTs), voice coil motors, and stepper motors—as well as the $\epsilon(t)$ on a millisecond basis throughout the night. These \SI{1}{\milli\second} telemetry streams are recorded by the CHARA messaging infrastructure for real-time diagnostics and post-processing of science data. This telemetry can be correlated with beam combiner data to support data quality assessment and reduction. Figures \ref{fig_target_spikes}, \ref{fig_target_spikes_fixed}, \ref{fig_met_error}, \ref{fig_step_response}, and \ref{fig_TransferFunction} are based on telemetry sampled at \SI{5}{\kilo\hertz}.

\subsection{Daily Maintenance}
Daily maintenance includes collecting telemetry files at the beginning of each night and verifying that the tracking residuals are within the expected range below \SI{12}{\nano\metre}. This procedure ensures system stability and optimal performance for high-precision interferometric measurements.

\begin{table}[ht]
\centering
\caption{Summary of Early Commissioning Issues and Mitigations}
\begin{tabular}{p{4.5cm} p{5.2cm} p{5.3cm}}
\toprule
\textbf{Issue (ID)} & \textbf{Root Cause} & \textbf{Resolution} \\
\midrule
\multicolumn{3}{c}{Unexpected Issues in the New System} \\
\midrule
1. Spurious and high-speed target position spikes caused cart tracking error 
(Fixed 2025 Jan) & Jitter in the time-tick counter of the metrology FPGA caused erroneous delay target generation at \SI{1}{\kilo\hertz}, resulting in target position spikes. The amplitude of these spikes was proportional to the cart speed. & The target position spikes are corrected by fixing the time-tick jitter in the metrology FPGA system. The cart tracking error reduced to below \SI{12}{\nano\metre}. \\
\midrule
2. Clock out-of-sync error caused target position spikes 
(Fixed 2025 Jun) & Clock synchronization between FPGA metrology systems and OPLE computer occasionally fails at power cycle; sync command ineffective unless issued post-reboot & Temporary workaround involves rebooting FPGA metrology hardware followed by the synchronization command \\
\midrule
\multicolumn{3}{c}{Expected} \\
\midrule
3. Lower control bandwidth 
(Fixed several times after 2024) & PID gains were initially tuned conservatively to avoid reacting to spurious target position spikes & Reoptimized PZT and voice coil PID loops after resolving timing errors to restore full bandwidth \\
\midrule
\multicolumn{3}{c}{Legacy Issues Also Present in the Old System} \\
\midrule
4. Loss of fringe visibility on the S1 telescope 
(Fixed 2023 Oct) & Mechanical vibrations from cable puller power supply mounted on beam transport vacuum tube & Relocated power supply of the vacuum infrastructure to eliminate vibration coupling \\
\bottomrule
\end{tabular}
\label{tab:operational_issues}
\end{table}

%%%%%%%%%%%%%%%%%%%%%%%%%
\section{Commissioning Results and Mitigations}\label{sec_initial_results}

Before the upgraded delay line system reached stable operations, several performance issues were encountered during early commissioning. These included both unexpected bugs in the newly developed hardware/software as well as issues previously present in the legacy system. Table~\ref{tab:operational_issues} summarizes key issues and their resolutions. The first two issues—target position spikes and limited bandwidth—were unexpected and unique to the new system. The remaining two were inherited from the legacy infrastructure but were also addressed during the commissioning process.

\subsection{Target Position Spikes Due to Time-Tick Jitter}

During early commissioning, diagnosing the root cause required substantial effort because the embedded metrology firmware is timing-critical and documentation of the delivered implementation was limited.

During initial testing, a systematic cart tracking error of $\epsilon(t)\sim\SI{80}{\nano\metre}$ RMS was observed (see Fig.~\ref{fig_target_spikes}). This error originated in the target generator within the embedded metrology system and was traced to time-tick jitter in the internal \SI{5}{\kilo\hertz} loop (see Sec.~\ref{sec_architecture} for the timing definition). In brief, occasional tick-count discontinuities-each 5 ~kHz tick should advance the 40~kHz counter by exactly 8 counts, but it occasionally did not—produced erroneous predictive delay targets proportional to the cart velocity, resulting in $\epsilon(t)\sim\SIrange{80}{100}{\nano\metre}$ RMS tracking errors and, for SPICA visible-light observations, fringe visibility losses exceeding 50\%.

A software workaround was implemented by rounding each \SI{25}{\micro\second} tick value to the nearest multiple of 8. This restored synchronization between the \SI{5}{\kilo\hertz} control loop and the \SI{40}{\kilo\hertz} clock, effectively eliminating the target-position spikes. As discussed in Sec.~\ref{sec_results}, this correction  significantly improved delay line tracking stability and tracking precision (see~Fig.~\ref{fig_target_spikes_fixed}).

Following the fixing of the target position spike issue, we optimized the PID servo control parameters for both the PZT actuators and the voice coil servos. This tuning improved system responsiveness, stability, and fringe tracking accuracy under both laboratory and on-sky conditions.

\subsection{Gain Optimization and Bandwidth Recovery}

During the early commissioning phase, prior to identifying the cause of the target position spikes, the PID gains, particularly for the PZT actuators, were deliberately reduced. This was done to prevent the PZT from overreacting to the spikes in the target positions. However, the unintended consequence was a large reduction in system bandwidth, which limited the ability to apply fast fringe tracker corrections.

After resolving the time-tick jitter issue in the metrology FPGA, we retuned the servo gains. The adjusted settings restored the desired system bandwidth, thereby enabling the high-speed, high-precision tracking required for real-time interferometric operations.

\begin{figure}
\centering
\includegraphics[width=\textwidth]{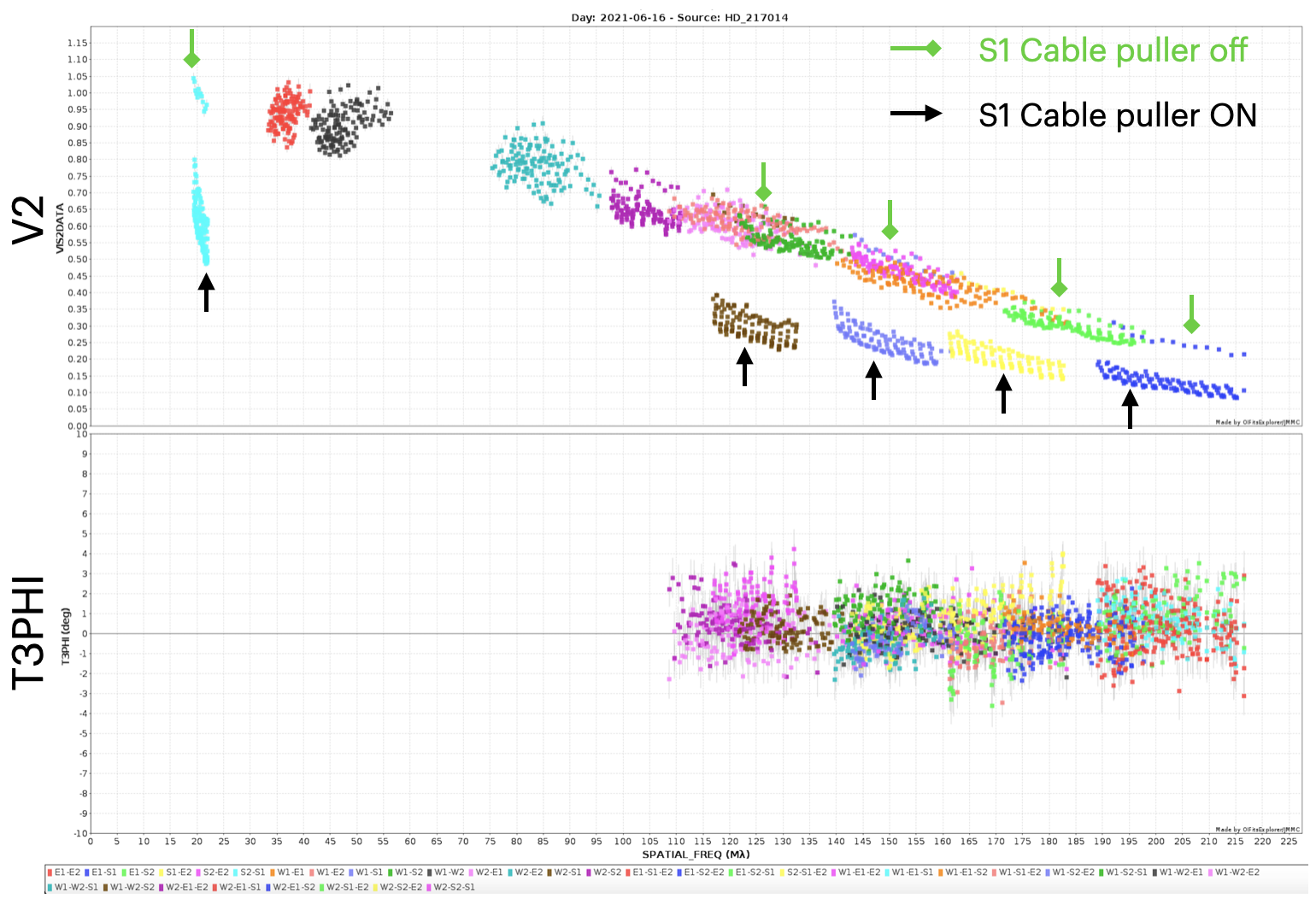}
\caption 
{ \label{fig_s1_v2_problem} 
Instrument visibility drop for the baselines associated with S1 telescope. Higher squared visibilities were recorded when the cable puller power supply was turned off (while the cart was moving forward), while significantly lower visibilities were observed when the power supply was active (while the cart was moving backwards). The visibility degradation is attributed to vibrations introduced by the power supply when mounted on the beam transport vacuum tube, which adversely affected optical path stability and fringe contrast. 
 } 
\end{figure}

\begin{figure}
\centering
\includegraphics[width=\textwidth]{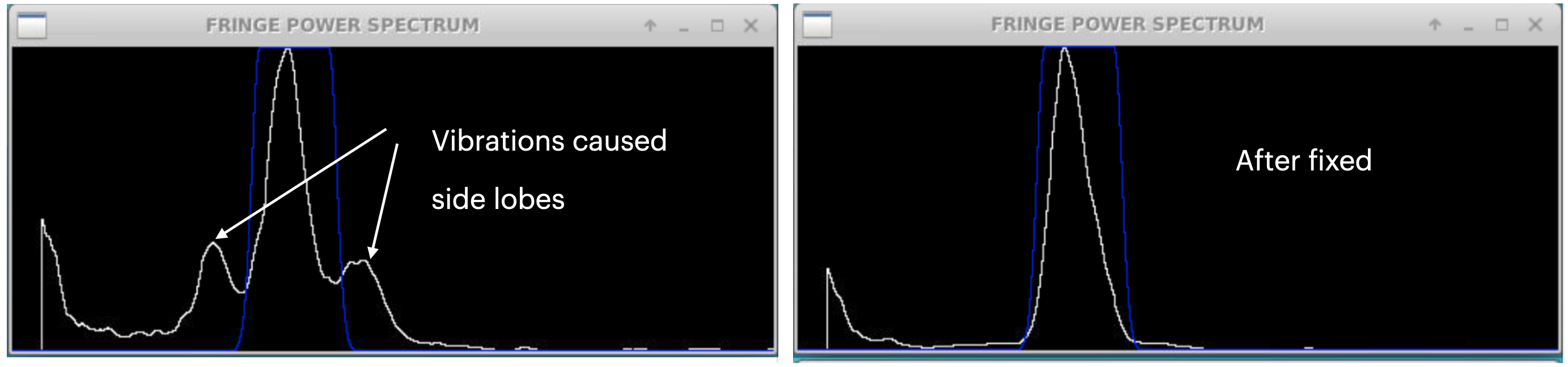}
\caption 
{ \label{fig_s1_v2_problem_fixed} 
CLASSIC instrument real-time power spectrum plots showing vibration issues caused by the S1 cable puller power supply. Left: Power spectrum observed with the CLASSIC instrument when the cable puller power supply was mounted on the beam transport vacuum tube. The presence of side peaks indicates mechanical vibrations, which contributed to a loss of visibility on the S1 telescope. Right: Power spectrum after relocating the power supply to the ground, eliminating the side peaks and resulting in a clean fringe peak. This confirms that mechanical coupling of the power supply to the vacuum tube was the source of the induced vibrations.
}
\end{figure} 

\subsection{S1 Visibility Loss Due to Cable Puller Vibrations}

In a separate vibration-related issue, a persistent reduction in fringe visibility was observed on the S1 telescope but only in specific regions of the sky (see Fig.~\ref{fig_s1_v2_problem}). This issue impacted data collected from roughly June 2021 through October 2023 for targets rising in the east (elevation $<55^\circ$ and azimuth $\sim$ 100$^\circ$). Using S1 as the reference cart mitigated the problem. After investigation, the issue was traced to mechanical vibrations introduced by the cable puller power supply. When the S1 delay cart moved in the backward direction, power to the cable puller was on. 

The power supply, which had been accidentally mounted on the beam transport vacuum tube, was found to induce mechanical vibrations that propagated through the POP mirror structure and degraded optical quality. These vibrations were confirmed via CLASSIC\cite{tenBrummelaar2013} instrument fringe power spectrum diagnostics (see Fig.~\ref{fig_s1_v2_problem_fixed}). Mechanical isolation of the power supply removed the vibration-induced side lobes in the power spectrum, indicating mechanical coupling rather than electrical interference. This mitigation led to a measurable improvement in fringe contrast for the S1 beam.

\begin{figure}
\centering
\includegraphics[width=\textwidth]{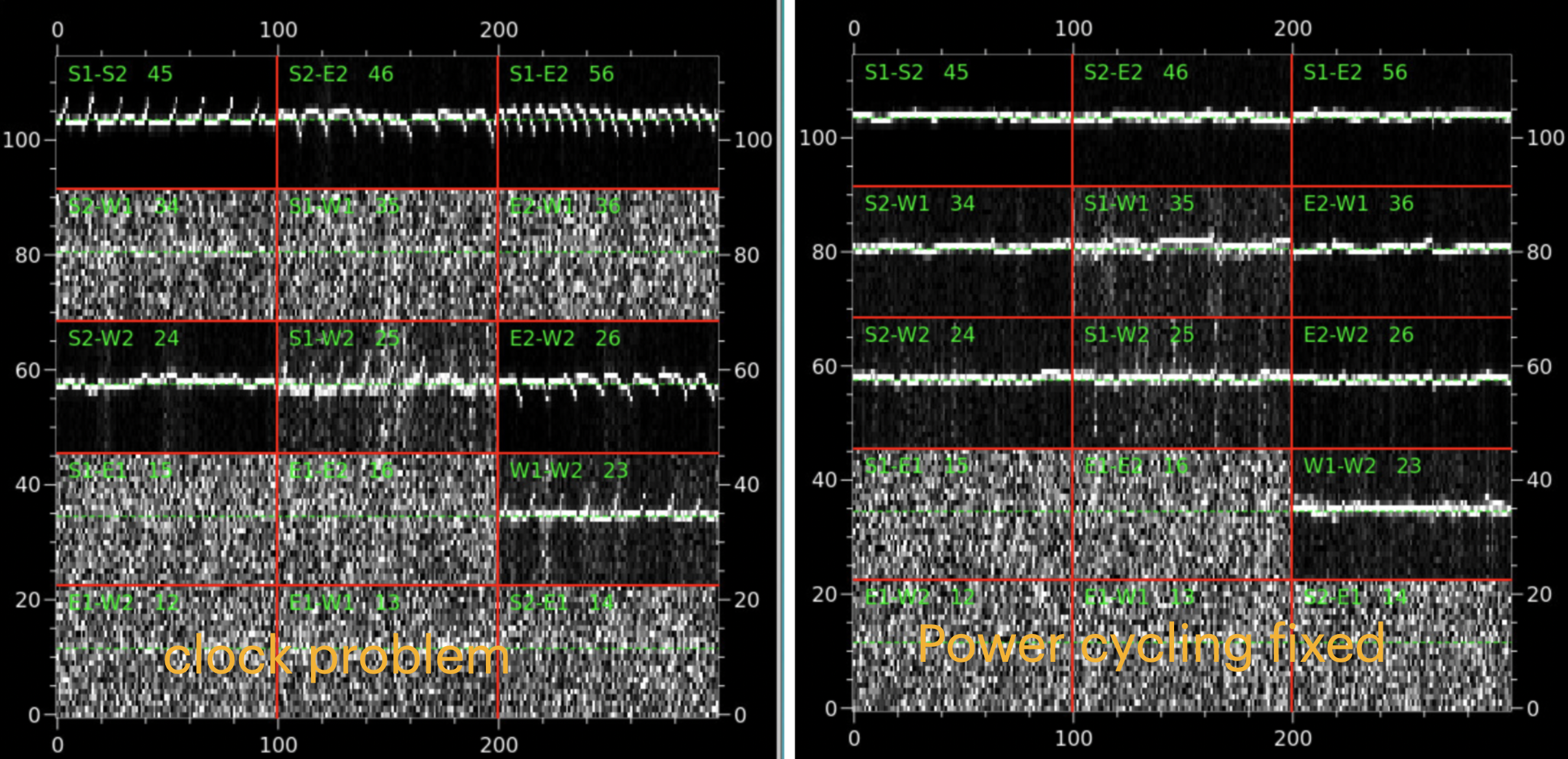}
\caption 
{ \label{fig_bootup_issue}
(Left): MYSTIC waterfall plot recorded during a clock synchronization issue between the \texttt{ople server} and the metrology FPGA systems. Periodic OPD spikes appeared across multiple baselines, caused by incorrect delay line positions when the \texttt{ople server} updated the baseline solution every six seconds. Due to misaligned clocks, the metrology system received offset corrections based on incorrect timing, introducing transient position errors. These appeared as sharp spikes followed by rapid corrections in the residual fringe tracking OPD.
(Right): After power cycling the metrology FPGA systems (S1 and E2), the synchronization issue was resolved and the residuals stabilized. Each of the 15 panels represents one baseline; the E1 telescope was not in use during this observation, so no fringe signals appeared for baselines associated with E1 telescope.
 } 
\end{figure} 

\subsection{Clock Synchronization Error between metrology FPGA systems and OPLE computer}

We also encountered an intermittent boot-time clock synchronization error between the metrology FPGA systems and the OPLE computer. This condition occurred approximately once every two weeks at random intervals. During routine nightly operations, the delay line control system is powered on at the start of the observing session, and all components typically initialize without issue. However, on rare occasions, the metrology clock fails to synchronize correctly, leading to optical path deviations as large as $\pm$\SI{400}{\nano\metre} peak-to-peak—much higher than the nominal RMS tracking error of $\sim$\SI{20}{\nano\metre}.

Fig.~\ref{fig_bootup_issue} shows an example of this anomaly, as revealed by large fringe tracking errors in the MYSTIC ``waterfall" plots. The waterfall plot shows the amplitude and position of the Fourier transform of a scan through the interferogram over time. Telemetry from the metrology system during such events displays spurious spikes in the laser position signal, indicating a transient instability in the clock synchronization. This issue is distinct from previously identified target timing spikes, as the delay line target remains static during the event.

In routine operations, this condition is detected at the start of the night by checking standard metrology diagnostics (e.g., tracking-error RMS over a short telemetry segment as in Fig.~\ref{fig_target_spikes_fixed}); if the RMS exceeds \SI{15}{\nano\metre}, the operator power-cycles the affected metrology FPGA, restarts \texttt{OPLECtrlQt5}, and then reissues the clock-synchronization command. For operational convenience, the diagnostic plot and the actions are executed remotely through the \texttt{OPLECtrlQt5} and \texttt{opletab} GUIs.

\subsection{Considerations for Data Reduction and Interpretation}

We recommend exercising caution when analyzing datasets acquired between mid-2021 and the end of 2024. Telemetry datasets (see Sec.~\ref{sec_telemetry}) can be used to crosscheck the science datasets. Tracking errors $\epsilon(t)$ in the delay lines can bias fringe visibility measurements between the target and calibrator. In severe cases, uncorrected or poorly tracked delay excursions may lead to systematic errors in stellar diameter estimates, binary orbit determinations, and surface imaging reconstructions.

%%%%%%%%%%%%%%%%%%%%%%%%%%%%%---------------------------------------
\section{Final On-Sky Performance Characterization}\label{sec_results}

\begin{table}[ht]
\centering
\caption{Laboratory Performance Metrics for the Six CHARA Delay Lines.}
\begin{tabular}{lcccccc}
\toprule
\textbf{Metric} & \textbf{S1} & \textbf{S2} & \textbf{E1} & \textbf{E2} & \textbf{W1} & \textbf{W2} \\
\midrule
Cart tracking RMS [nm] & 9 & 5 & 11 & 12 & 12 & 10 \\
$-$3\,dB Bandwidth [Hz] & 125 & 130 & 105 & 100 & 110 & 106 \\
Correction Settling Time [ms] & 1 & 1 & 1 & 1 & 1 & 1 \\
Estimated $R$-band Visibility Loss [\%] & 1 & 1 & 5 & 3 & 1 & 1 \\
\bottomrule
\end{tabular}
\label{tab:lab_performance}
\end{table}

\begin{table}[ht]
\centering
\caption{Comparison of Legacy and Upgraded Delay Line Systems. }
\begin{tabular}{lcccc}
\toprule
Parameter & Legacy System & Upgraded System & VLTI & KI\cite{Colavita2013}\\
\midrule
Residual cart tracking RMS error, $\epsilon(t)$ [nm]& 10-20 & $<12$ & 20\cite{Derie2000} & 10-20 \\
Fringe tracker RMS residuals, $\Delta$OPD [nm] & 260\cite{Monnier2010,Monnier2012CHAMP}  & 90-230\cite{Pannetier2022} (Sec.~\ref{sec_onsky}) & 75-250\cite{Lacour2019} & 63-171\\
\bottomrule
\end{tabular}
\label{tab:comparison}
\end{table}

%----------------------------------
\subsection{Step response and power spectrum characterization}

\subsubsection{Step response, settling time}
To evaluate the performance of the CHARA delay line system, we analyzed residual tracking error (RMS) and system dynamics using time- and frequency-domain diagnostics. 

Table~\ref{tab:lab_performance} summarizes the cart tracking errors, all within \SI{12}{\nano\metre} RMS (see~Fig.~\ref{fig_target_spikes_fixed}), consistent with the performance achieved at other leading interferometric facilities such as PTI\cite{Colavita1999}, KI\cite{Colavita2013}, and VLTI \cite{Derie2000}. While the tracking error performance ($\epsilon(t) \leq\SI{12}{\nano\metre}$ RMS) is comparable across systems, a direct comparison of frequency response is not possible for the legacy CHARA system due to the lack of documented measurements. The dynamic behavior of the system was further characterized using step response analysis. By introducing a step offset change in the target position and observing the delay line response, we measured \SI{1}{\milli\second} response time it takes for the stage to settle within a defined tolerance band ($\pm5$\% of the final value).

\begin{figure}
\centering
\includegraphics[width=0.75\textwidth]{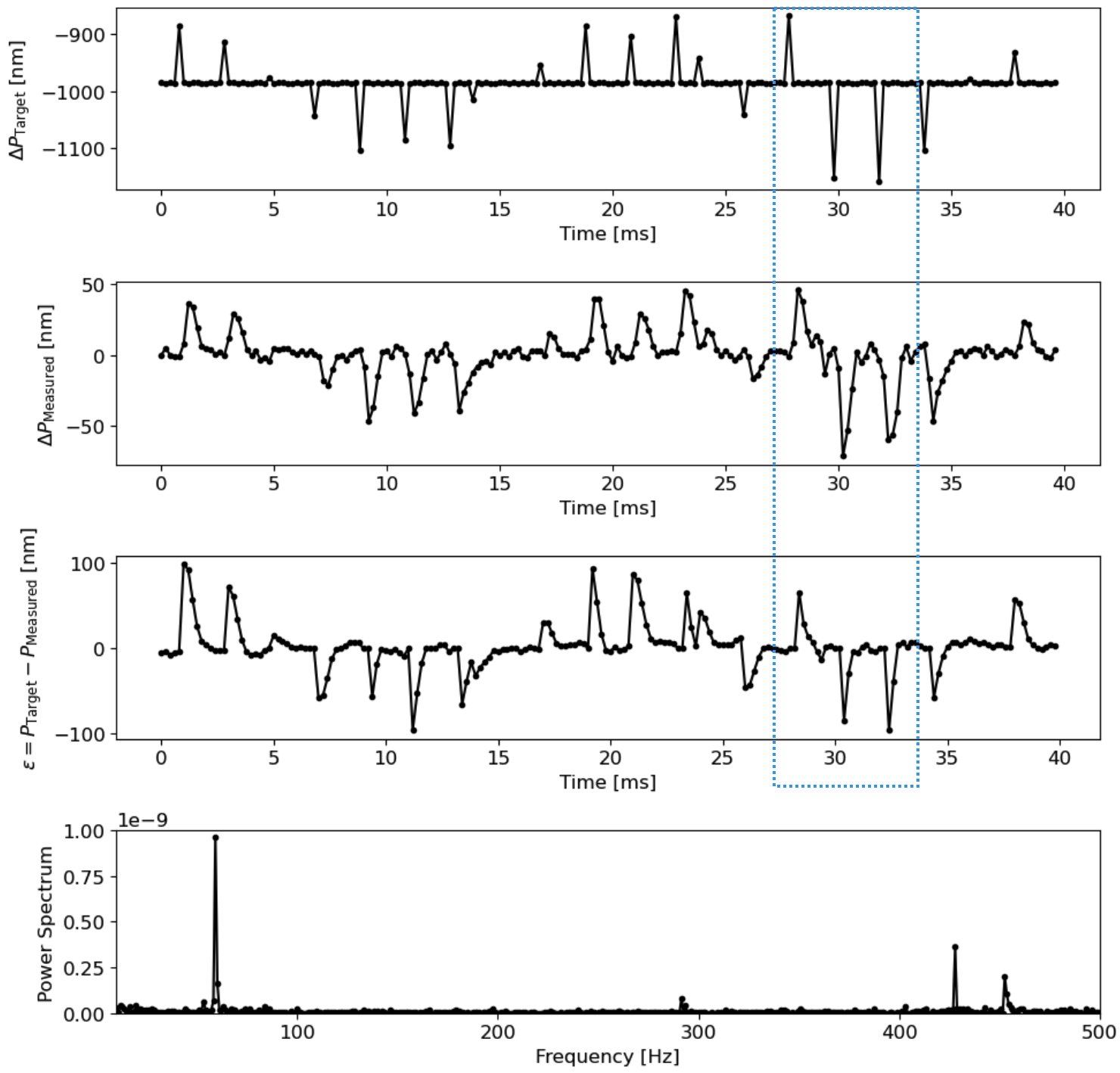}
\caption
{\label{fig_met_error}
On-sky delay-line tracking on 2026 Feb 05 with SPICA-FT phase tracking at \SI{500}{\hertz}. Top: target trajectory combining a baseline solution of \SI{-4.2}{\milli\metre\per\second} and superimposed \SIrange{50}{100}{\nano\metre} steps every \SI{2}{\milli\second}. Second: measured cart position. Third: tracking error $\epsilon(t)$, typically within \SI{12}{\nano\metre} RMS. Bottom: error PSD with a dominant \SI{60}{\hertz} component. Fig.~\ref{fig_step_response} shows the zoom-in of the dotted box data.}
\end{figure}

\begin{figure}
\centering
\includegraphics[width=0.75\textwidth]{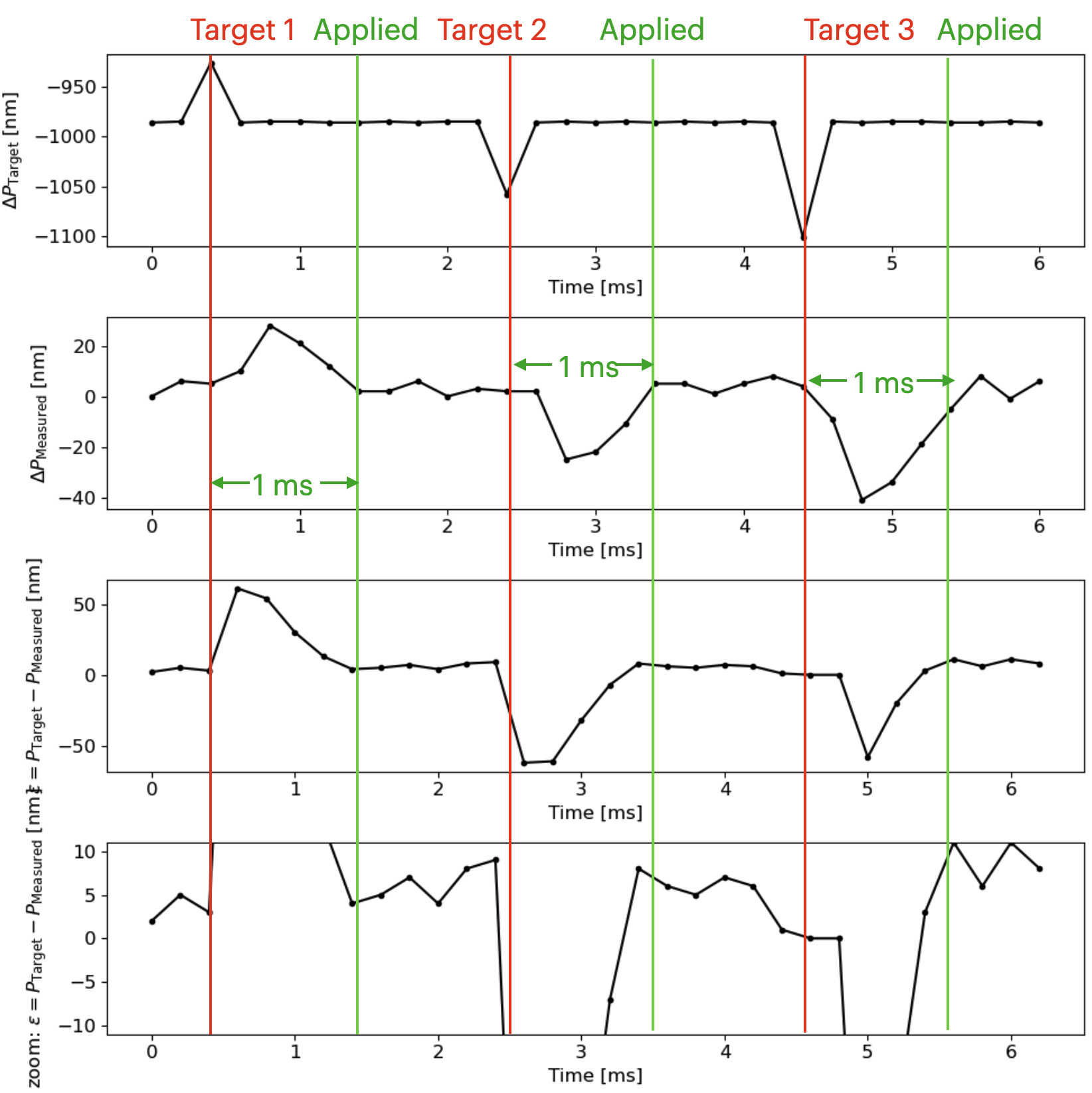}
\caption
{\label{fig_step_response}
Zoomed view of Fig.~\ref{fig_met_error} showing the step response over a few ms. The settling time (within $\pm$5\% of the final value) is $\sim\SI{1}{\milli\second}$, and the overshoot/ripple is within $\pm$\SI{10}{\nano\metre}. Here each data point is taken at 5~KHz sampling rate. }
\end{figure}

Figures~\ref{fig_met_error}  shows representative on-sky step-like OPD commands applied during SPICA-FT phase tracking, with an overview (40~ms).  The \SI{500}{\hertz} SPICA-FT OPD offset update rate is at the limit of the delay-line end-to-end latency of $\sim$\SI{2}{\milli\second} (Sec.~\ref{sec_puredelay}). A typical OPD offsets are within $\pm0.5~\mu$m. A zoom-in of step response is shown in Fig.~\ref{fig_step_response}. It indicates a settling time of $\sim$\SI{1}{\milli\second} (within $\pm$5\%) and a small, lightly damped ripple with peak-to-peak amplitude $\lesssim$\SI{10}{\nano\metre}. Faster settling can be achieved by increasing loop gains, at the cost of larger overshoot and ringing; we therefore tune the loops to balance response time and damping. For larger commanded offset steps $>\SI{0.5}{\micro\metre}$, the overshoot/ripple (proportional to step amplitude) is larger than our specification RMS error. For larger group-delay offsets, typically in micron-scale, used in packet tracking and fringe search scanning, offsets are therefore implemented as a sequence of smaller increments (\SI{0.5}{\micro\metre}) to keep the actuators in their linear range and to limit overshoot.

\begin{figure}
\centering
\includegraphics[width=\textwidth]{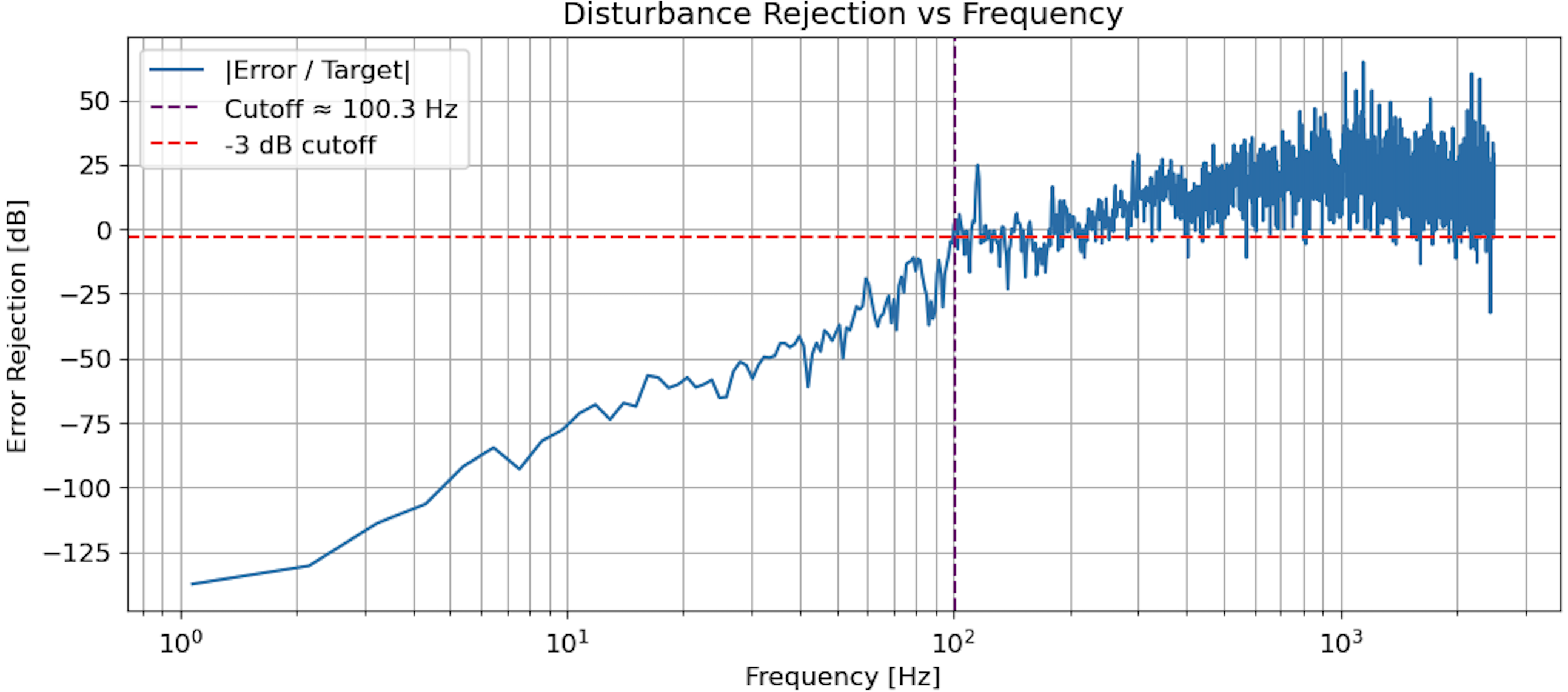}
\caption 
{\label{fig_TransferFunction}
Transfer function analysis of the CHARA delay line system for telescope E1. The figure presents the frequency-domain characterization of the delay line control system, while the cart was tracking at \SI{6.54}{\milli\metre\per\second} using the baseline solution and phase tracking with SPICA-FT on 2025 Jun 14. The measured $-3$dB rejection cutoff frequency is approximately \SI{100}{\hertz}, marked by the dashed magenta line.
}
\end{figure}

\subsubsection{Power spectrum analysis}
We further analyzed the frequency characteristics of the delay line tracking errors using power spectral density (PSD) analysis. This approach provides insight into residual mechanical vibrations and periodic disturbances that are not fully corrected by the servo control system. The PSD revealed prominent peaks at \SI{60}{\hertz} and harmonics, consistent with resonant modes associated with the US power line frequency (see Fig.~\ref{fig_met_error}). 

Although various servo gain settings were tested to mitigate these resonances, the peaks could not be fully suppressed. Further investigation is needed to identify the underlying sources of these persistent disturbances. Possible contributors include mechanical vibrations from vacuum pumps used to maintain pressure in the beam transport tubes, as well as aging components in the delay line infrastructure that may introduce resonances during cart motion. Despite the presence of these residuals, the cart tracking error remains within \SI{12}{\nano\metre}, satisfying the technical performance requirements of the CHARA delay line system.

\subsubsection{Transfer Function and Pure Delay}\label{sec_puredelay}
Finally, we computed the system transfer function to assess its frequency response. This analysis involves calculating the complex gain and phase shift between the target and position signals in the frequency domain. 

Fig.~\ref{fig_TransferFunction} shows the $-3$dB cutoff frequency for E1 telescope cart, that is approximately \SI{100}{\hertz}. Table~\ref{tab:lab_performance} summarizes the $-3$dB cutoff frequency for all the six carts, ranging from \SIrange{100}{130}{\hertz}. The cutoff frequency is limited due to the pure delay caused by the control loop system. We measured a maximum pure delay of up to \SI{2}{\milli\second} between the time the fringe tracker transmits an OPD correction and the moment that correction is fully realized by the delay line system. This latency arises from three primary sequential stages in the control loop. First, once the fringe tracker determines a correction, it sends the command to the metrology system over a dedicated local low-latency Ethernet connection. This transmission process takes $\sim\SI{0.25}{\milli\second}$. Upon receiving the command, the metrology system requires on average \SI{0.5}{\milli\second} time to update the target position within its \SI{1}{\kilo\hertz} target update loop. Finally, the PZT takes on average \SI{1}{\milli\second} to actuate the correction with settling time. Altogether, this end-to-end response time of $\leq$\SI{2}{\milli\second} defines the system latency, which is well within acceptable limits for real-time fringe tracking applications at CHARA. 

SPICA-FT servo control goals are operating at \SIrange{250}{500}{\hertz}. At SPICA-FT OPD offset servo control rates approaching \SI{500}{\hertz}, this 2~ms delay line latency becomes a limiting factor because it starts to introduce a phase lag. We are pursuing a two-step reduction of the end-to-end latency toward $\sim$\SI{1.5}{\milli\second}: (i) a faster target-position update within \SI{0.25}{\milli\second} in metrology FPGA, and (ii) gain optimization to achieve PZT settling within \SI{0.75}{\milli\second}. These changes have not yet been validated on sky.

To compare with VLTI, \SIrange{2.2}{2.8}{\milli\second} level latency delays are reported  when fringe-tracking corrections are applied through the GRAVITY tip-tilt-piston (TTP) stage and the common-path VLTI main delay lines  \cite{Woillez2024,Laugier2024GravityPlus}. However, as part of the GRAVITY+ upgrade, low latency differential delay lines in the beam-compressor train (BCDDL) are being deployed in the VLTI recombination laboratory; their reduced pure delay is intended to extend high-frequency disturbance rejection control (notably in the 30--100~Hz range), and their control reuses elements of the PRIMA delay-line hardware \cite{Pepe2008ESPRI,Laugier2024GravityPlus}.

\begin{figure}
\centering
\includegraphics[width=\textwidth]{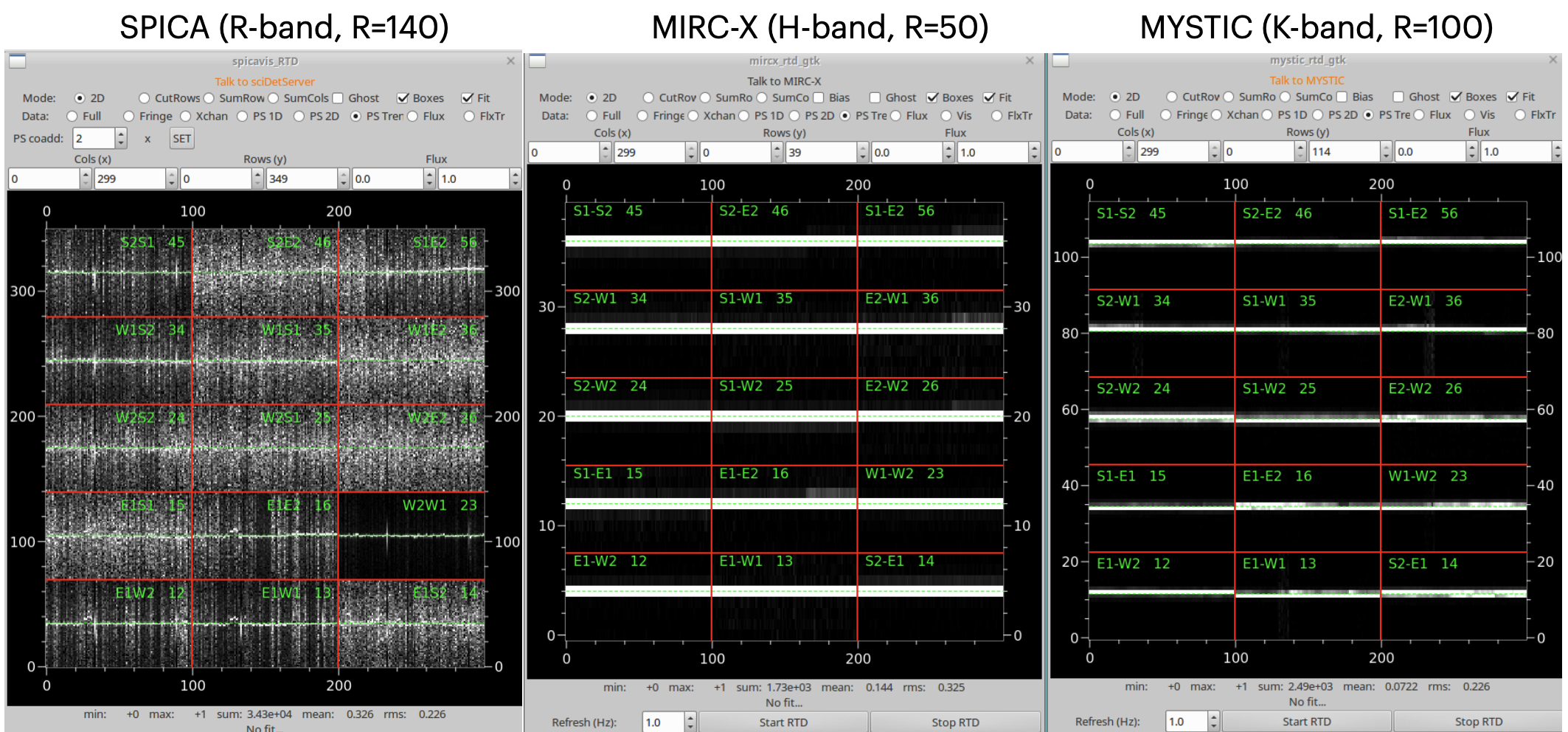}
\caption 
{ \label{fig_spica_mircx_mystic}
An example observation from UT 2025 July 08 is shown, displaying waterfall plots for SPICA (left), MIRC-X (middle), and MYSTIC (right). Each sub panel shows the OPDs for all 15 baselines of the respective instrument, with beam pairings indicated in the top-right corner of each subplot (e.g., 12 = Telescope 1–2). Each subplot shows the temporal evolution of the OPD residuals along the vertical axis, with time increasing along the horizontal axis. 
Brighter regions represent higher fringe contrast, corresponding to successful fringe detection. During this observation, SPICA-FT and MIRC-X ($H$-band) were fringe tracking, while SPICA and MYSTIC simultaneously recorded data. The coherent integration times were \SI{8}{\milli\second} for MIRC-X, and \SI{20}{\milli\second} for both MYSTIC and SPICA. The spectral resolutions used were $\mathcal{R}=50$ for MIRC-X, $100$ for MYSTIC, and $240$ for SPICA. 
A zoomed time-series view of fringe-tracker residuals is provided in Fig.~\ref{fig_mircx_pd}.
}
\end{figure}

\begin{figure}[ht]
\centering
\includegraphics[width=0.95\textwidth]{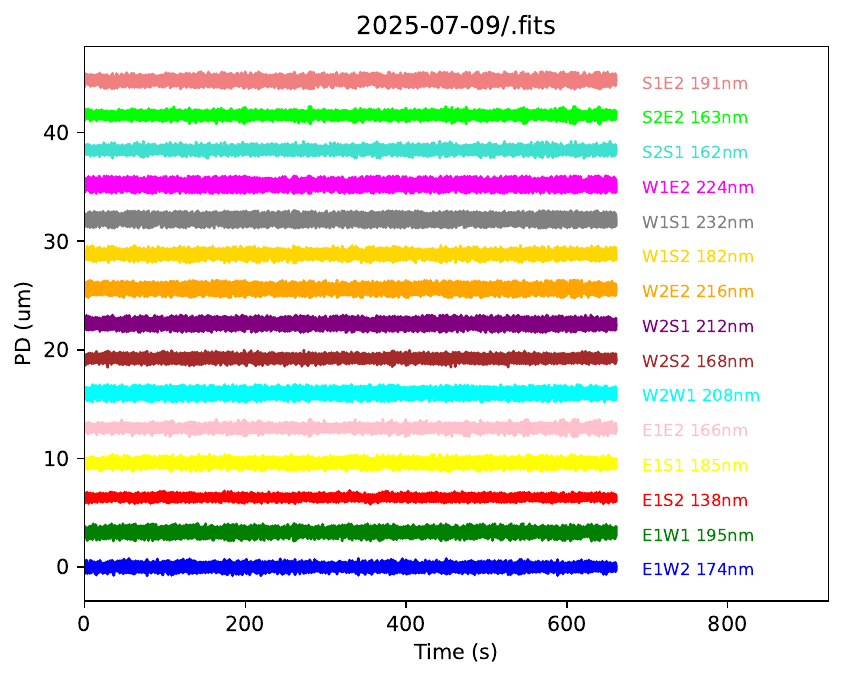}
\caption 
{ \label{fig_mircx_pd}
An example SPICA-FT observation recorded on UT 2025~June~14. This shows residual OPDs for all 15 baselines of the CHARA Array. The OPDs are maintained near zero with \SIrange{140}{230}{\nano\metre} RMS. These residuals are sampled at the SPICA-FT frame rate (\SI{250}{\hertz} for these data). } 
\end{figure}

Finally, the delay-line tracking error ($\epsilon(t)\lesssim\SI{12}{\nano\metre}$ RMS; Fig.~\ref{fig_met_error}) quantifies how accurately the delay lines follow their commanded targets, whereas the fringe-tracker residuals in Fig.~\ref{fig_spica_mircx_mystic} and ~\ref{fig_mircx_pd} represent the residual atmospheric piston after sensing and correction. As a result, fringe-tracker residuals include atmospheric turbulence, structural vibrations, fringe-tracker measurement noise, and control strategy in addition to the delay-line contribution. Further reduction in phase tracking residuals requires a combined approach: higher fringe-tracker frame rates and/or predictive control (e.g., Kalman filtering as implemented for GRAVITY; \cite{Lacour2019}), vibration mitigation, and reduced end-to-end latency in the delay-line correction path.

As summarized in Table~\ref{tab:lab_performance}, the system is well-optimized for $-3$dB rejection cutoff frequency below \SI{100}{\hertz}. Differences in measured bandwidths for carts with similar RMS tracking errors (e.g., E1, W1, W2) primarily reflect cart-specific mechanical dynamics (mass loading, cable drag, and rail conditions) and conservative loop tuning to avoid resonances.

However, $R$-band observations demand faster corrections, corresponding to a coherence time of $\tau_0 = \SI{6}{\milli\second}$. A control bandwidth in the range of \SIrange{100}{130}{\hertz}—equivalent to $\tau_0 = \SIrange{7.5}{10}{\milli\second}$—is adequate to suppress the majority of lower-frequency atmospheric piston variations, which carry most of the turbulence power. A detailed analysis of the system transfer function is presented in Appendix~\ref{sect_TF_model}.

Table~\ref{tab:comparison} presents a comparison of fringe tracking residuals achieved with the upgraded CHARA delay line system, alongside those reported for the legacy CHARA system, the VLTI, and the Keck Interferometer (KI). The results demonstrate that SPICA-FT (operating in the $H$-band) achieves improved performance relative to the legacy CHARA system, which was based on the $K$-band CHAMP instrument~\cite{Monnier2010,Monnier2012CHAMP}. SPICA-FT benefits from operating at a shorter wavelength, allowing for more precise phase sensing compared to $K$-band instruments.

\begin{figure}
\centering
\includegraphics[width=0.85\textwidth]{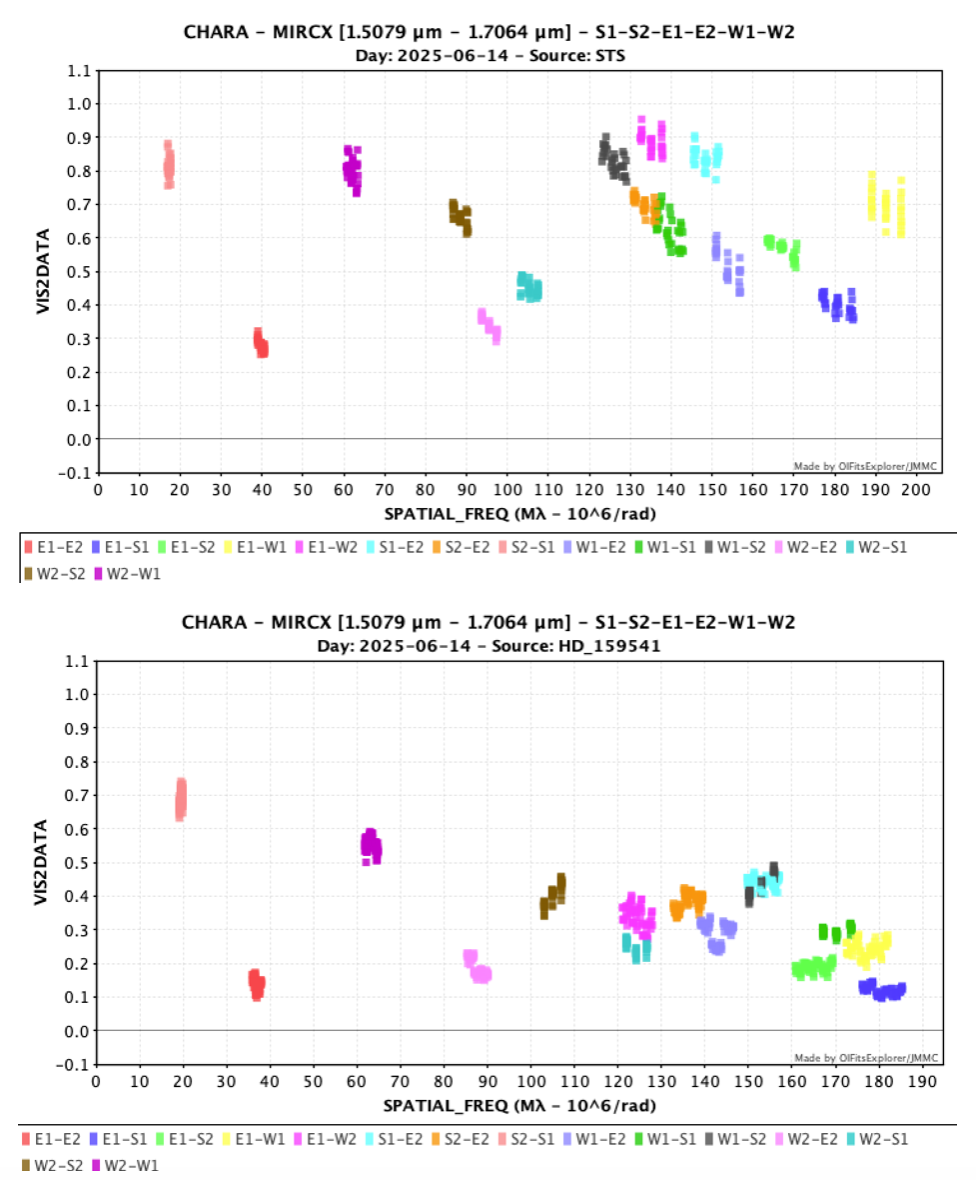}
\caption 
{ \label{System_visibility} 
Raw (uncalibrated) system visibility of MIRC-X in the $H$-band.
The top panel shows the squared visibilities measured with the STS internal laboratory source, which provides a collimated point-like beam. The STS is located downstream of the delay lines, so the light bypasses atmospheric turbulence, delay line tracking errors, polarization mismatches, and wavelength dispersion—isolating the instrumental response of the MIRC-X beam combiner itself. The beam order is E1–W2–W1–S2–S1–E2; in this configuration, baselines formed by physically adjacent beams are expected to exhibit higher visibility, consistent with the MIRC-X fringe sampling design.
The bottom panel presents on-sky raw visibility measurements. Compared to the STS results, a noticeable reduction in visibility is observed. This degradation primarily arises from residual atmospheric aberrations not fully corrected by adaptive optics and fringe tracking systems. There is also a small reduction in visibility with increasing spatial frequency because of the stellar angular diameter of $\sim$ 0.44 mas.
 } 
\end{figure}

\begin{figure}
\centering
\includegraphics[width=0.92\textwidth]{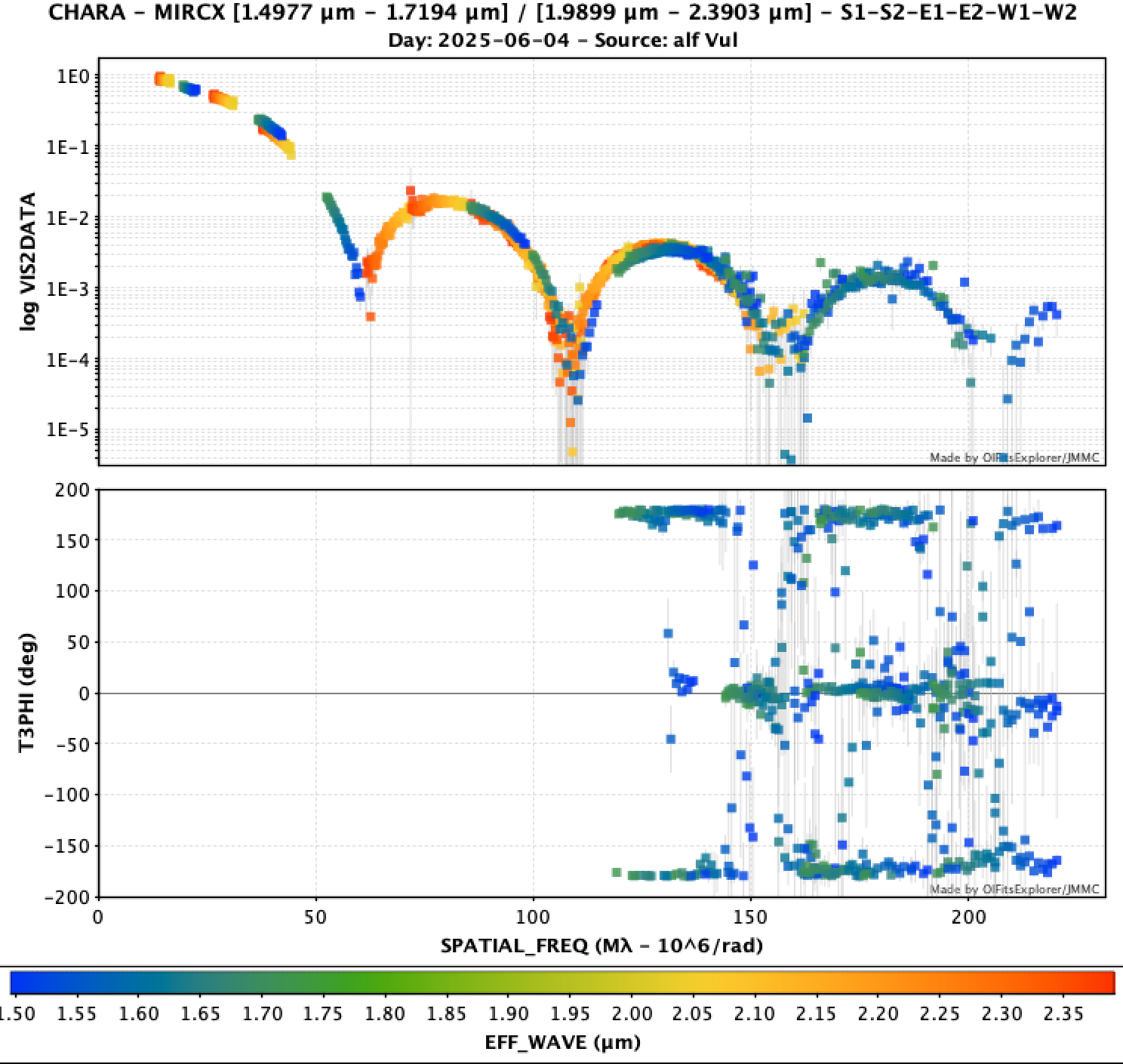}
\caption 
{ \label{fig_v2_t3phi} 
Squared visibility (top panel) and closure phase (bottom panel) observables were recorded for $\alpha$~Vul on UT 2025~June~4, using both MIRC-X and MYSTIC instruments in the $H$- and $K$-band wavelengths. The closure phase is an observable quantity for a triangle
of baselines. The calibrated data reveal higher-contrast fringes extending through at least four visibility nulls, demonstrating the excellent performance of the upgraded delay line system. 
 } 
\end{figure}

Laboratory and on-sky testing indicate that the CHARA delay lines contribute $\leq2$\% to fringe contrast degradation at $R$-band, with the largest cart tracking error, with an RMS residual $\epsilon(t)\leq\SI{12}{\nano\metre}$. For observations in the $H$ and $K$-bands, the delay line contribution to visibility loss remains below 1\%, satisfying the operational performance criteria. These results are consistent with the technical requirements outlined in Sec.~\ref{sec_tech_require}.

%----------------------------------
\subsection{On-sky fringe visibility characterization}\label{sec_onsky}

The on-sky performance of the upgraded CHARA delay lines is assessed through measurements of fringe visibility and its stability across science and calibrator observations. For a standard observation night, the $R$-, $H$-, and $K$-band beam combiners are co-phased on a laboratory Six Telescope Simulator (STS)\cite{Anugu2020} before on-sky observations. The fringe tracking during on-sky science observations is typically performed in the $H$-band using SPICA-FT, while data is recorded in the $R$-band (SPICA) or $K$-band (MYSTIC. SPICA-FT is a software application implemented on the MIRC-X hardware and software platform; SPICA taps the visible channel upstream of the H/K dichroic, while MIRC-X and MYSTIC share the near-IR path after the dichroic, so NCOPD terms arise primarily after the beam split (Fig.~\ref{fig_spica_mircx_mystic_schematic}).
Since atmospheric turbulence evolves more rapidly at shorter wavelengths --- scaling as $\propto\lambda^{6/5}$ in both the Fried parameter $r_0$ and coherence time $\tau_0$ --- this leads to uncorrected residual OPD fluctuations in visible-light observations. Additionally, the SPICA-FT ($H$-band) and SPICA ($R$-band) beams do not share the same optical path, introducing further NCOPD errors, mainly wavelength dispersion.  While SPICA includes internal delay lines to manually correct these slow drifts ($< 1$ Hz), it does not support self-fringe tracking. In the case of MYSTIC ($K$-band), its observations benefit from slower atmospheric phase fluctuations relative to the SPICA-FT fringe tracker. SPICA-FT and MYSTIC share a common optical path, reducing NCOPD errors. Although differential water vapor columns between $H$ and $K$-bands introduce slow $\Delta$OPD drifts, these are measured by MYSTIC beam combiner and corrected in real-time without significant loss of fringe contrast.

Following the fixing of the delay line issues outlined in Sec.~\ref{sec_initial_results}, CHARA now routinely achieves simultaneous on-sky fringe observations (see Fig.~\ref{fig_spica_mircx_mystic}) with SPICA ($R$-band), MIRC-X ($H$-band), and MYSTIC ($K$-band). Fig.~\ref{fig_spica_mircx_mystic} displays representative group delay tracking waterfall plots with no fringe losses or cycle slips, illustrating stable tracking. Fig.~\ref{fig_mircx_pd} shows the OPD residuals measured during SPICA-FT operations in phase-tracking mode. The observed residuals of $\Delta$OPD $> \SI{140}{\nano\metre}$ RMS are slightly higher than those reported for the VLTI/GRAVITY fringe tracker\cite{Lacour2019}. SPICA-FT has also demonstrated best-case residuals of $\sim$\SI{90}{\nano\metre} RMS under favorable seeing conditions (SPICA-FT team, private communication); these results are not shown here and will be presented in the dedicated fringe-tracker paper. SPICA-FT with servo control at \SI{500}{\hertz} is being tested. Fig.~\ref{fig_step_response} shows a response of delay lines when OPD offsets commanded at \SI{500}{\hertz} rate and the cart error is smaller within $\pm10$ nm.

A full fringe-tracker error analysis (including PSDs, atmospheric piston, vibrations, measurement noise, control strategy and mitigation strategies) is being led by the SPICA-FT team (D. Mourard); this manuscript focuses on delay-line performance.
End-to-end, operating SPICA-FT at \SI{500}{\hertz} with a \SI{2}{\milli\second} detector integration time, combined with the $\sim$\SI{2}{\milli\second} delay-line correction latency, implies a sensing-to-actuation delay of order \SI{4}{\milli\second}. This is just below the representative $R$-band coherence time $\tau_0\approx\SI{6}{\milli\second}$ (Table~\ref{tab:tracking_require}), so \SI{500}{\hertz} operation is plausible but with limited margin under average conditions. Under good seeing, $\tau_0$ can approach \SI{11}{\milli\second} in $R$-band (Table~\ref{tab:tracking_require}), in which case this latency is not expected to be the dominant limitation.

Fig.~\ref{System_visibility} compares system visibilities observed with MIRC-X in the $H$-band using both the STS internal laboratory point source and on-sky observations of a calibrator star. The STS data characterize the intrinsic visibility performance of the instrument in the lab, while the on-sky visibilities show an additional 10–15\% degradation across all 15 baselines, attributed to residual atmospheric effects, polarization mismatches, and chromatic dispersion. MYSTIC exhibits a similar level of visibility degradation in the $K$-band. A detailed analysis of SPICA system visibilities is currently in preparation by Mourard et al. 

Fig.~\ref{fig_v2_t3phi} presents calibrated squared visibility measurements for a representative science target with an angular diameter of 4 mas. The data clearly resolve four visibility nulls across the CHARA baselines, illustrating high-contrast fringe detection and exceptional angular resolution. A dedicated astrophysical analysis of these data, including limb-darkening measurements for $\sim$three dozen stars observed after the delay-line fixes, is in preparation (Anugu et al., in prep.). To our knowledge, CHARA data of this quality --- showing such a high number of well-resolved visibility nulls --- has not been previously published.

%%%%%%%%%%%%%%%%
\section{Summary and Future Applications}\label{sec_future}

From mid-2021 through the end of 2024, the CHARA Array delay lines underwent a modernization to replace the legacy control system that had operated reliably for over two decades but had begun to show signs of aging. While the upgrade was not intended to improve the fundamental tracking performance (RMS error and bandwidth), it improved maintainability and fault isolation, and enabled systematic telemetry logging for operational diagnostics and science data interpretation. The modular architecture also simplifies future expansion of the delay line system if additional telescopes are added to the array (Sec.~\ref{sec_architecture}). As part of this effort, we carried out a characterization of the delay line system that was not available for the legacy implementation.

Key performance issues were identified during initial operations and resolved later, including time-tick-jitter-induced target spikes, control gain optimization, and other visibility loss issues (see Sec.~\ref{sec_initial_results}). Post-upgrade, the system achieves accurate fringe tracking: cart tracking errors are consistently maintained within $\epsilon(t) \leq \SI{12}{\nano\metre}$ RMS, with a measured $-3$~dB bandwidth of \SIrange{100}{130}{\hertz} (Sec.~\ref{sec_results}). The fringe-tracker residuals, $\Delta$OPD $= \SIrange{90}{230}{\nano\metre}$, are better than or comparable to the legacy system (Table~\ref{tab:comparison}). This level of performance satisfies the requirements (Sec.~\ref{sec_tech_require}) for high-precision interferometry across the $R$, $H$, and $K$-bands.
Further reduction of phase residuals will require identifying and filtering optical-path vibrations, as demonstrated for GRAVITY \cite{Lacour2019,Laugier2024GravityPlus}.

We discuss applications and requirements for upcoming projects at CHARA.

\subsection{Sky coverage and CHARA Michelson Array Pathfinder (CMAP)}

The CHARA Michelson Array Pathfinder (CMAP)\cite{Koehler2024,Scott2024} project aims to extend the maximum baseline of the Array to \SI{1}{\kilo\metre} to increase angular resolution at CHARA.

CMAP observations may face two primary challenges: (i) the delay line carts must operate at higher speeds---up to \SI{20}{\milli\metre\per\second}---to accommodate longer baselines while using the existing \SI{46}{\metre} physical delay infrastructure; and (ii) the effective sky coverage or observing window may be reduced due to the limited \SI{92}{\metre} total optical delay available for a \SI{1}{\kilo\metre} baseline. Northern targets with declinations above $65^\circ$ are already limited to observations with only five telescopes at the current maximum baseline of \SI{331}{\metre}, showing the constraints imposed by the existing delay line stroke range. At the sidereal delay rate of $\sim$\SI{8}{\milli\metre\per\second}, the \SI{92}{\metre} cart stroke corresponds to $\sim$\SI{3.2}{\hour} of continuous tracking before a POP change is required; in practice, POP switches are scheduled between target sequences.

Following the resolution of time-tick jitter and gain stability issues, the upgraded system now supports cart speeds up to \SI{20}{\milli\metre\per\second} without loss of tracking performance.
 
\begin{table}[ht]
\centering
\caption{CHARA Array baseline lengths and position angles for telescope pairs, together with the available fixed-delay POP settings (Pipes of Pan). POP delays are implemented as switchable, discrete optical path segments (typically in \SI{36.6}{\metre} increments) that extend the geometric delay budget and keep the main carts within range during tracking.}
\begin{tabular}{l llll}
\toprule
\textbf{Tel} & \textbf{Pairing} & \textbf{Length (m)} & \textbf{Position Angle (°)} \\
\midrule
S1 & E1 & 330.66 & 22.28  \\
  & E2 & 278.76 & 14.63  \\
  & W1 & 278.50 & 321.02  \\
  & W2 & 210.97 & 340.88  \\
  & S2 & 34.07 & 350.29  \\
\midrule
S2 & E1 & 302.33 & 25.70  \\
  & E2 & 248.13 & 17.87  \\
  & W1 & 249.39 & 317.18  \\
  & W2 & 177.45 & 339.08  \\
\midrule
E1 & W1 & 313.53 & 253.39  \\
  & W2 & 221.82 & 241.27  \\
  & E2 & 65.88 & 236.60  \\
\midrule
E2 & W1 & 251.34 & 257.73  \\
  & W2 & 156.27 & 243.23 \\
\midrule
W1 & W2 & 107.92 & 99.11  \\
\midrule
POP (fixed delay) & 1 & 0.00  & ---  \\
   & 2 & 36.60 & ---   \\
   & 3 & 73.20 & ---  \\
   & 4 & 109.70 & ---  \\
   & 5 & 143.10 & ---  \\
\bottomrule
\end{tabular}
\label{tab:chara_grouped}
\end{table}

To further enhance sky coverage, a flexible reallocation of the existing POP assignments can be pursued \cite{Ridgway1994Pipes,Bagnuolo1996OPLE,Sturmann2021Pipes}. POPs provide switchable fixed-delay segments (typically in \SI{36.6}{\metre} increments; Table~\ref{tab:chara_grouped}) that offload large, quasi-static geometric delays so that the main carts remain within their dynamic range for fine tracking (see Sec.~\ref{sec_overview_delaylines}). In the current configuration for SPICA, MIRC-X, and MYSTIC, the W2 beamline is fixed at POP~5; redistributing the remaining POP settings across other beamlines can increase accessible sky for long baselines, including CMAP.

An alternative approach involves extending the delay lines without adding new POPs, through a double-pass configuration utilizing: (i) in the existing \SI{46}{\metre} physical cart travel, and/or (ii) in the current vacuum POP pipes\cite{Ridgway1994Pipes,Bagnuolo1996OPLE}. 

The first variant of this solution would double the usable optical delay by sending the beam through the delay line twice. However, this introduces several drawbacks, including the need for larger optics, customized OPLE cart designs, and additional skew reflections to alternate between horizontal and vertical planes, potentially introducing polarization artifacts. Additionally, doubling the physical travel range to \SI{92}{\metre} complicates cable management and increases the required atmospheric dispersion compensation due to the longer air path, which may reach up to \SI{331}{\metre} in some array configurations.

The second solution of implementing dual-pass in POPs could increase the maximum POP delay to \SI{246}{\metre} from the current single pass \SI{143}{\metre}, but at the cost of two additional mirror reflections and associated throughput losses.

\subsubsection{Dual-Field Interferometry and Long Exposures for Off-Axis Targets}
To achieve higher sensitivity interferometric observations, dual-field interferometry has been successfully employed\cite{Shao1992, Eisenhauer2017}, where two widely separated astronomical targets are observed simultaneously using separate beam combiners. One beam combiner functions as a phase tracker, locking onto a bright reference star to correct atmospheric turbulence OPD variations in real time. The second beam combiner takes long exposures on a faint science target, benefiting from the stabilized fringes provided by the phase-tracking channel.

In the CHARA context, this mode achieved first-light observations in 2025 using the $\alpha$~Piscum binary system (HD~12447 and HD~12446, mag $H/K\sim3.5$) with MIRC-X as the fringe tracker in the $H$-band and MYSTIC as the science instrument in the $K$-band (Anugu et al., submitted). The projected maximum dual-field separation is $\sim$5~arcsec \cite{Anugu2025SPIE}. With the delay-line issues resolved and SPICA-FT demonstrating $\sim$90~nm best-case phase-tracking residuals, dual-field observations can be extended to fainter companions by taking longer exposures on the science beam combiner.

A future option to mitigate the $\sim$\SI{2}{\milli\second} end-to-end pure-delay term is to add a dedicated fast piston actuator (e.g., a short-stroke PZT stage) in the beam-combiner laboratory on the MIRC-X + MYSTIC table and close the high-bandwidth fringe-tracking loop on this actuator. The actuator slowly varying mean position would then be offloaded to the existing CHARA delay-line carts to keep the fast stage within its dynamic range. In principle, reducing the effective actuation latency in this way could enable SPICA-FT operation at $\sim$\SI{1}{\kilo\hertz} and further reduce residual OPD, improving coherent long-exposure observations of faint off-axis targets with MYSTIC.

\subsubsection{Spectro-interferometry}
Spectro-interferometry enables spatially and spectrally resolved studies of astrophysical objects, probing accretion, mass loss, disk kinematics, exoplanet atmospheres, and spin-orbit alignment in multiple systems. This capability is critical for understanding stellar evolution and planet formation at high resolution.

MIRC-X ($\mathcal{R}$=6000) and MYSTIC ($\mathcal{R}$=1724) support these science cases by offering high-resolution interferometric spectroscopy. To operate near the instrument background limit with long exposures, both instruments rely on stable phase tracking, enabled by the upgraded delay line system.

\subsubsection{High-Contrast Nulling Interferometry}
High-contrast nulling interferometry is a promising future direction for CHARA, aiming to suppress bright starlight and detect faint nearby structures such as exoplanets, brown dwarfs, and exozodiacal dust \cite{Bracewell1978}. The technique achieves destructive interference of on-axis stellar light by introducing a $\pi$ phase shift, thereby enhancing sensitivity to off-axis companions. Nulling interferometry requires highly stable flux and phase control over the duration of the null. The demonstrated performance of the CHARA delay line system and ongoing improvements in phase tracking suggest feasibility for future nulling experiments at the CHARA Array.
Using the standard small-phase-error relation for a two-beam nuller \cite{Bracewell1978}, $N_{\phi}\simeq (\pi \sigma_{\mathrm{OPD}}/\lambda)^2$, an OPD residual of the range $\sigma_{\mathrm{OPD}}=\SIrange{90}{230}{\nano\metre}$ corresponds to $N_{\phi}\sim(1.7$--$11)\times10^{-2}$. These estimates represent phase leakage only; achieving deeper nulls also requires controlling intensity mismatch, polarization, chromatic dispersion, and instrumental systematics. Multi-night group-delay observations at CHARA have achieved detection contrasts down to $\sim2\times10^{-4}$ \cite{Gardner2026}. This contrast metric is not directly comparable to null depth, but it provides a useful benchmark for CHARA high-contrast performance; nulling offers a complementary path by suppressing stellar leakage within a single observing sequence, with ultimate limits set by systematic leakage terms beyond phase noise.

\section*{Disclosures}
The authors declare there are no financial interests, commercial affiliations, or other potential conflicts of interest that have influenced the objectivity of this research or the writing of this paper.

\section*{Code, Data, and Materials Availability} 
The data used in this paper are available through the CHARA Open Online Database Portal: \url{https://www.chara.gsu.edu/observers/database}. The MIRC-X and MYSTIC datasets presented in Figs.~\ref{fig_s1_v2_problem}, \ref{System_visibility}, and \ref{fig_v2_t3phi} were reduced using the mircx\_pipeline, which is publicly accessible at \url{https://gitlab.chara.gsu.edu/lebouquj/mircx_pipeline.git}. These figures were plotted using the Jean-Marie Mariotti Center OIFITS Explorer tool, available at \url{https://amhra.jmmc.fr/}.

\section* {Acknowledgments}
We thank the anonymous referees for their constructive suggestions, which improved the clarity and quality of this paper.
The CHARA delay line control system upgrade was designed and delivered by AZ Embedded Systems. We thank Tim Buschmann and Brad Hines for their valuable contributions to the project. We also thank the JPL team whose pioneering delay-line architecture enabled the legacy system. We acknowledge the CHARA Array technical staff: M. Anderson, K. Kubiak, R. Koehler, C. Lanthermann, H. Renteria, S. Ridgway, J. Sturmann, L. Sturmann, and C. Woods for their contributions to CHARA development, maintenance, and observational support. We thank the SPICA and MIRC-X collaborations for developing the instruments that enabled key aspects of this work. 
This work is based upon observations obtained with the Georgia State University Center for High Angular Resolution Astronomy Array at Mount Wilson Observatory. The CHARA Array is supported by the National Science Foundation under Grant No.\ AST-1636624, AST-2034336, and AST-2407956. Institutional support has been provided from the GSU College of Arts and Sciences and the GSU Office of the Vice President for Research and Economic Development.
JDM acknowledges funding for the development of MIRC-X (NASA-XRP NNX16AD43G, NSF-AST 2009489) and MYSTIC (NSF-ATI 1506540, NSF-AST 1909165). This research utilized the Aspro and SearchCal services of the Jean-Marie Mariotti Center.
SK acknowledges funding for MIRC-X received funding from the European Research Council (ERC) under the Horizon 2020 research and innovation programme of the European Union (Starting Grant No. 639889 and Consolidated Grant No. 101003096). 
This project has received funding from the European Research Council (ERC) under the Horizon 2020 research and innovation programme of the European Union (Grant agreement No. 101019653) as part of the SPICA instrument developmenet and science. This research has made use of the Jean-Marie Mariotti Center - AMHRA service at https://amhra.jmmc.fr/.

%%%%% References %%%%%

\bibliography{report}  % bibliography data in report.bib
\bibliographystyle{spiejour}  % makes bibtex use spiejour.bst

\appendix
\makeatletter
% Avoid duplicate hyperref anchors when section numbering resets in the appendix.
\renewcommand{\theHsection}{app.\Alph{section}}
\makeatother

\section{Acronyms}\label{sec:acronyms}

\begin{multicols}{2}
\begin{acronym}
 \acro{AO}{Adaptive Optics}
 \acro{CHARA}{Center for High Angular Resolution Astronomy}
 \acro{CHARIOT}{CHARA Array integrated optics testbench}
 \acro{CMAP}{CHARA Michelson Array Pathfinder}
 \acro{CPU}{Central Processing Unit}
 \acro{DC}{Direct Current}
 \acro{FPGA}{Field-Programmable Gate Array}
 \acro{GPS}{Global Positioning System}
 \acro{GUI}{Graphical User Interface}
 \acro{HI-5}{High-contrast Interferometry up to 5 \textmu m}
 \acro{JPL}{Jet Propulsion Laboratory}
 \acro{KI}{Keck Interferometer}
 \acro{LTS}{Long-Term Support}
 \acro{MIRC-X}{Michigan InfraRed Combiner-eXeter}
 \acro{MYSTIC}{Michigan Young Stellar Imager at CHARA}
 \acro{NCOPD}{non-common path optical path difference}
 \acro{NIOS}{Nios Embedded Processor}
 \acro{NTP}{Network Time Protocol}
 \acro{OPD}{Optical Path Difference}
 \acro{OPLE}{Optical Path Length Equalizer}
 \acro{PA}{Position Angle}
 \acro{PC}{Personal Computer}
 \acro{PCI}{Peripheral Component Interconnect}
 \acro{PD}{Proportional-Derivative}
 \acro{PID}{Proportional-Integral-Derivative}
 \acro{PLL}{Phase-Locked Loop}
 \acro{POP}{Pipes of Pan}
 \acro{PSD}{Power Spectral Density}
 \acro{PTI}{Palomar Testbed Interferometer}
 \acro{PZT}{Piezoelectric Transducer}
 \acro{RMS}{Root Mean Square}
 \acro{SPICA}{Stellar Parameters and Images with a Cophased Array}
 \acro{SPICA-FT}{SPICA Fringe Tracker}
 \acro{STS}{Six Telescope Simulator in the lab}
 \acro{UDP}{User Datagram Protocol} 
 \acro{UTC}{Coordinated Universal Time}
 \acro{VC}{Voice Coil}
 \acro{VLTI}{Very Large Telescope Interferometer}
 \acro{VME}{Virtual Machine Environment}
\end{acronym}
\end{multicols}

\section{Transfer function model}
\label{sect_TF_model}

The dynamic behavior of the OPLE cart system is governed by the combined effects of its mechanical structure, actuators, and amplifiers. This collective response can be modeled using a transfer function, which provides a concise representation of the input-output relationship of the system in the Laplace domain. Conceptually, a transfer function encapsulates the differential equations governing the system dynamics and is typically expressed as a ratio of polynomials in the complex frequency variable \( s \).

For the OPLE control system, experimental system identification and tuning analysis have revealed that the dominant response characteristics can be modeled using a transfer function that includes two zeros and one pole, corresponding to a standard PID-like control structure. The system transfer function can be expressed as:

\[
H(s) = K \cdot \frac{(s - z_1)(s - z_2)}{(s - p_1)}
\]

\noindent where:
\begin{itemize}
  \item \( K \) is the overall system gain,
  \item \( z_1 \) and \( z_2 \) are the system zeros introduced by the derivative and proportional actions,
  \item \( p_1 \) is the dominant system pole, typically associated with integrator dynamics or plant lag.
\end{itemize}

This structure shapes the system response (phase margin and overshoot) while maintaining tracking stability. Accurate modeling of this transfer function supports servo loop tuning, disturbance rejection, and stable delay line operation at higher cart speeds required for long-baseline interferometry.

%%%%% Biographies of authors %%%%%
\vspace{2ex}\noindent\textbf{Narsireddy Anugu} is an optical scientist at the CHARA Array, Georgia State University. He received his PhD in Physics Engineering from the University of Porto, Portugal, and subsequently held postdoctoral fellowships at the University of Exeter, the University of Michigan, and the University of Arizona. His research interests include high-angular resolution astronomy, interferometry, and imaging techniques.

%\listoffigures
%\listoftables

\end{document}